\documentclass[useAMS,usenatbib]{mn2e}

\pdfoutput=1

\usepackage[pdftex]{graphicx}
\usepackage[fleqn]{amsmath}

\newcommand{\mpl}{M_{\mathrm p}}

\newcommand{\rp}{r_{\mathrm p}}

\newcommand{\op}{\Omega_\mathrm{p}}
\newcommand{\okep}{\Omega_\mathrm{K}}
\newcommand{\cs}{c_\mathrm{s}}
\newcommand{\me}{\mathrm{M_\oplus}}
\newcommand{\xs}{x_\mathrm{s}}

\newcommand{\be}{ \begin {equation}}
\newcommand{\ee}{ \end {equation}}

\newcommand{\sigp}{\Sigma_\mathrm{p}}

\title{A torque formula for non-isothermal Type I planetary migration - I. Unsaturated horseshoe drag}
\author[S.-J. Paardekooper, C. Baruteau, A. Crida and W. Kley]{S.-J. Paardekooper$^{1}$\thanks{E-mail:
S.Paardekooper@damtp.cam.ac.uk}, C. Baruteau$^{2}$, A. Crida$^{1,3}$ and W. Kley$^{3}$\\
$^{1}$DAMTP, University of Cambridge, Wilberforce Road, Cambridge CB3 0WA, United Kingdom\\
$^{2}$UCO/Lick Observatory, UC Santa Cruz, 1156 High Street, Santa Cruz, CA 95064, USA\\
$^{3}$Institut f\"ur Astronomie \& Astrophysik, Universit\"at T\"ubingen, Auf der Morgenstelle 10, 72076 T\"ubingen, Germany}
\begin{document}

\date{Draft version \today}

\pagerange{\pageref{firstpage}--\pageref{lastpage}} \pubyear{2008}

\maketitle

\label{firstpage}

\begin{abstract}
We study the torque on low-mass planets embedded in protoplanetary discs in the two-dimensional approximation, incorporating non-isothermal effects. We couple linear estimates of the Lindblad (or wave) torque to a simple, but non-linear, model of adiabatic corotation torques (or horseshoe drag), resulting in a simple formula that governs Type I migration in non-isothermal discs. This formula should apply in optically thick regions of the disc, where viscous and thermal diffusion act to keep the horseshoe drag unsaturated. We check this formula against numerical hydrodynamical simulations, using three independent numerical methods, and find good agreement. 
\end{abstract}

\begin{keywords}
planetary systems: formation -- planets and satellites: formation.
\end{keywords}


\section{Introduction}
Planets are thought to form in circumstellar discs around young stars. In the core accretion model, gas giant planets emerge in the disc through gas accretion onto a previously accumulated solid core of a few times the mass of the Earth \citep[$\me$,][]{pollack96}. An alternative scenario involves direct fragmentation of the disc \citep{boss97}, which is probably only possible in the outermost regions of the disc \citep{boley09}.

In general, objects embedded in protoplanetary discs will exchange angular momentum with the disc, which leads to a change in their orbital parameters. The nature of this interaction depends on the masses of the object and the disc. Small bodies, up to a few km in size, on a Keplerian orbit will experience a head wind from the gas, since the gas is partially supported by pressure and will thus orbit at sub-Keplerian velocity \citep{weiden77}. This head wind will lead to orbital decay, the time scale of which can be as short as a few $100$ yrs \citep{weiden77}.

The most massive objects, approximately the mass of Jupiter, can tidally truncate the disc, forming a deep annular gap around their orbits \citep{linpap86II}. The planet, being repelled by both gap edges, is locked inside the gap and will slowly accrete onto the central star with the rest of the disc \citep{linpap86III}. The minimum mass for this Type II migration to occur depends on the scale height and viscosity of the disc \citep{crida06}.

Planets that are not massive enough to open up a gap, but do significantly perturb the disc, can be subject to a very rapid mode of migration called Type III when embedded in a very massive disc \citep{maspap03}. The mechanism of Type III migration relies on a distortion of streamlines in the coorbital region due to a radial flow of gas with respect to the planet \citep{adamthesis}. This radial flow of gas can be due to the migration of the planet itself, resulting in a positive feedback with the possibility of a runaway process \citep{maspap03}, with migration time scales of the order of a few tens of dynamical time scales. Sustaining this rapid mode of migration has proved to be very difficult \citep{adamin,adamout}.

In this paper, we will be concerned with planets that do not significantly perturb the disc, which is typically valid for objects up to a few $\me$. This regime of Type I migration was long thought to be the simplest case, since it could be treated using a linear analysis. It was shown in \cite{gt79} that the torque exerted on the planet by the disc can be decomposed in a wave torque, arising at Lindblad resonances, and a corotation torque, generated at corotation resonances. This analysis was subsequently refined \citep{art93,ward97}, eventually resulting in a semi-analytical torque formula for isothermal discs \citep{tanaka02}. This formula has been confirmed by fully non-linear, isothermal, hydrodynamical calculations \citep{bate03,dangelo3D}. 

The time scale for Type I migration is inversely proportional to the mass of the planet and the disc, but is typically $10^{4-5}$ yr for a $1$ $\me$ planet embedded in a Minimum Mass Solar Nebula \citep{ward97,tanaka02}. This is worrying, since the lifetime of the disc is of the order of $10^{6-7}$ yr, making the survival of low-mass planets highly unlikely. Planetary synthesis models have great difficulties reproducing the observed semi-major axis distribution when including Type I migration, and need to reduce the Type I torque from an order of magnitude \citep{ida08} to as much as a factor of $1000$ \citep{alibert05,mordasini09}. 

Above results were obtained without considering magnetic effects. It has been shown that including magnetic fields, either regular \citep{terquem03}, or turbulent \citep{nelson04}, may slow down or even stop Type I migration. It is not clear, however, if protoplanetary discs are sufficiently ionised throughout to couple effectively to the magnetic field. 

There is another important ingredient missing in standard models of Type I migration, which is to release the isothermal assumption and account for the energy balance in a more realistic way. A growing body of studies is dedicated to this problem, dealing with high-mass planets \citep{dangelo03,klahr06}, shadowing-effects \citep{jang05}, and opacity jumps \citep{menou04}. In \cite{paard06}, it was shown through three-dimensional, radiation-hydrodynamic simulations that for deeply embedded low-mass planets, Type I migration could be qualitatively different from the isothermal case. Planets could suddenly move outward as well as inward, depending on the local opacity. This result was confirmed using two-dimensional simulations with a self-consistent heating and cooling balance \citep{kley08}.  
It was subsequently shown that this effect was due to the effect of a radial entropy gradient in the disc on the corotation torque \citep{baruteau08,paard08}, and non-linear in nature \citep{paardpap08}. 

In this series of papers, we aim at catching the essential physics of the non-linear, non-isothermal corotation torque in a simple model that can be used to predict the Type I migration rate, as a function of radial density and temperature gradients. In this paper, we consider the unsaturated, adiabatic horseshoe drag, combined with a linear estimate for the wave torque. Effects of viscous and thermal diffusion will be considered in a forthcoming work. We start in section \ref{secEq} with reviewing the basic equations and disc models, and describe our numerical methods in section \ref{secNum}. We give a more detailed overview of isothermal Type I migration in section \ref{secTypeI}. In section \ref{secModel} we present a simple model for the torque on a low-mass planet in the presence of both entropy and vortensity gradients, and subsequently compare this model to numerical simulations in section \ref{secTest}. A short discussion is given in section \ref{secDisc}, and we present our conclusions  in section \ref{secCon}.


\section{Basic equations}
\label{secEq}
\subsection{Governing equations}
The basic equations are those of the conservation of mass, momentum and energy for a two dimensional disc in a  frame rotating with angular velocity $\op.$ We adopt a cylindrical polar coordinate system $(r,\varphi)$ with the origin $(r=0)$ located at the central mass. The continuity equation and the equation of motion take  the form 
\begin{equation}
\frac{\partial \Sigma}{ \partial t}= -\nabla\cdot(\Sigma  {\bf v})
\label{eqcont}
\end{equation}
and 
\begin{equation} 
\frac{D {\bf v}}{D t} +2\op{\hat {\bf k}} \times {\bf v}
=-\frac{1}{\Sigma }\nabla p - \nabla\Phi 
\label {eqmot}
\end{equation}
respectively, while, in the adiabatic case, entropy is conserved along streamlines:
\begin{equation}
\frac{D (p/\Sigma^\gamma)}{D t}=0,
\label{eqent}
\end{equation}
where $\gamma$ is the adiabatic exponent. Above, $\Sigma$ denotes the surface density, ${\bf v}$ the velocity, $p$ is the vertically integrated pressure, $\Phi$ is the gravitational potential and ${\bf \hat k}$ is the unit vector in the vertical direction. The convective derivative  is defined by 
\be \frac{D}{Dt} \equiv  \frac{\partial}{\partial t}+ {\bf v}\cdot \nabla.\ee
In the remainder of this paper, we will refer to $s\equiv p/\Sigma^\gamma$ as the entropy of the fluid. An ideal gas equation of state was used, $p=\mathrm{R_g}\Sigma T/\mu$, where $\mathrm{R_g}$ is the gas constant, $\mu$ is the mean molecular weight and $T$ is the temperature. We neglect effects of self-gravity, viscosity and thermal diffusion. The potential $\Phi$ contains terms due to the central mass $M_*$, and direct and an indirect term due to the planet \citep[see][]{nelson00}. 

\subsection{Equilibrium models}
We construct axisymmetric equilibrium models that have power law profiles in surface density and temperature, with indices $-\alpha$ and $-\beta$ respectively. This means that the initial entropy profile is a power law as well, with index $-\xi$, where
\begin{equation}
\xi=\beta-(\gamma-1)\alpha.
\label{eqxi}
\end{equation}
The angular velocity is Keplerian, with a slight correction for the radial pressure gradient to maintain pressure equilibrium. The temperature at the location of the planet is chosen so that the pressure scale height at the location of the planet is $H_\mathrm{p}\equiv h\rp$, with $h\ll 1$. Typically we use $h=0.05$. In the absence of self-gravity, the density at the location of the planet $\sigp$ can be chosen arbitrarily. 

\subsection{Planet}
The potential of the planet, located at $r=\rp$ and $\varphi=\varphi_\mathrm{p}$, is taken to be a softened point mass:
\begin{equation}
\Phi_\mathrm{p}=-\frac{G\mpl}{\sqrt{r^2+\rp^2-2r\rp\cos(\varphi-\varphi_\mathrm{p})+b^2\rp^2}},
\end{equation}
with $b$ the softening parameter. In order to approximately account for 3D-effects, $b$ should be comparable to $h$. Typically we use $b=0.4h$. When calculating the torque on the planet, we include all disc material. We have checked that excluding a fraction of the Hill sphere in the torque calculation does not affect the results. Below, we will use $q$ to denote the mass ratio $\mpl/M_*$.

\section{Numerical methods}
\label{secNum}
Equations \ref{eqcont}, \ref{eqmot} and \ref{eqent} are solved on a cylindrical grid, extending from $r/\rp=0.4$ to $r/\rp=1.6$, and the full $2\pi$ in azimuth. The typical resolution amounts to $\Delta r/\rp=0.0013$ and $\Delta \varphi=0.0025$. Due to the small radial extent of the horseshoe region, a large radial resolution is required. We have checked that taking square cells around the planet's location (by doubling the resolution in $\varphi$) does not influence the results. 

We have used three independent numerical codes: RODEO \citep[ROe solver for Disc Embedded Objects,][]{rodeo}, based on an approximate Riemann solver, FARGO \citep[Fast Advection in Rotating Gaseous Objects,][]{fargo,fargo2}, and RH2D \citep[Radiation Hydrodynamics in 2 Dimensions,][]{kley89,kley99}. The latter two methods are based on the van Leer upwind algorithm.

RODEO is based on the general relativistic Roe solver outlined in \cite{eulderink95}. It uses stationary extrapolation to integrate gravitational and geometrical source terms, and can handle arbitrary coordinate frames. Since it is based on a Riemann solver, RODEO is specifically designed to handle sharp discontinuities, usually in the context of shocks. We will see in section \ref{secTest} that although shocks do not play a role for low-mass planets, discontinuities arise in the flow for which the use of a Riemann solver can be an advantage.

RH2D is a 2D mixed explicit/implicit second-order upwind algorithm that also uses a staggered grid. It can treat radiation transport in the flux-limited diffusion approximation, but in this paper we are only concerned with adiabatic discs. Its advection algorithm is based on the monotonic transport scheme by \cite{vanLeer77}. 

FARGO\footnote{\tt http://fargo.in2p3.fr/} is a 2D hydrodynamical code, using a polar grid centred on the star. It solves the Navier-Stokes and continuity equations, as well as the energy equation in a more recent version \citep{baruteau08}, which we use here. It is based on van Leer upwind algorithm, on a staggered mesh. Both FARGO and RH2D use the FARGO algorithm\,: in each ring $i$, at every time-step, the averaged azimuthal velocity $\bar{v}_{\varphi,i}$ is computed. The ring is globally shifted by the corresponding number of cells for the considered time-step length $\delta t$\,:
$n_i=E[\bar{v}_{\varphi,i}\frac{\delta t}{r\delta \varphi}]$, where $\delta \varphi$ is the elementary angle associated to a cell, and $r$ the radius of the ring. Then, the advection is performed using the remnant azimuthal velocity in every cell $v'_\varphi=v_\varphi-n_i\,r\,\delta \varphi/\delta t$. In rotating disks where $|v'_\varphi| \ll\bar{v}_\varphi$, this enables a speed-up of the computation, and a lower numerical diffusivity becasue of the larger time step.

\section{Isothermal Type I migration}
\label{secTypeI}
In this section, we briefly review recent progress on Type I migration in the isothermal limit. This will prove helpful in understanding the general case, since very similar processes operate.

One can linearise equations \ref{eqcont} and \ref{eqmot} and solve these numerically using outgoing wave boundary conditions \citep{kory93}. This yields a Lindblad torque \citep[][their 2D result]{tanaka02}: 
\begin{equation}
\Gamma_\mathrm{L}/\Gamma_0=-3.2-1.468\alpha,
\label{eqTLindliniso}
\end{equation}   
with\footnote{All torques presented in this paper will be normalised by $\Gamma_0$. Note that $\Gamma_0$ is proportional to $q^2$.} 
\begin{equation}
\Gamma_0=(q/h)^2\sigp\rp^4\op^2,
\end{equation}
and a corotation torque
\begin{equation}
\Gamma_\mathrm{c,lin}/\Gamma_0=2.04-1.36\alpha.
\end{equation}   
Note that the linear corotation torque is proportional to the radial gradient of specific vorticity \citep{gt79}, being zero for $\alpha=3/2$. It is also important to stress that these results were obtained with essentially no gravitational softening. 

It was shown in \cite{drag} that corotation torques are non-linear in general, unless a very strong viscosity is applied. The linear corotation torque is replaced by non-linear horseshoe drag \citep{ward91}:
\begin{equation}
\Gamma_\mathrm{HS}/\Gamma_0=\frac{3}{4}\left(\frac{3}{2}-\alpha\right) \xs^4 \frac{h^2}{q^2},
\end{equation}
where $\xs$ is the half-width of the horseshoe region, in units of $\rp$. In the limit $b \rightarrow 0$, it was shown in \cite{horse} that $\xs^2=1.68q/h$, making the horseshoe drag
\begin{equation}
\Gamma_\mathrm{HS}/\Gamma_0=2.11\left(\frac{3}{2}-\alpha\right)=3.18-2.11\alpha,
\label{eqTHSiso}
\end{equation}
which is a factor of more than $3/2$ larger than the linear corotation torque. The result that the non-linear corotation torque is larger than its linear counterpart also holds for non-zero gravitational softening.

One can then combine equations \ref{eqTLindliniso} and \ref{eqTHSiso} to obtain a formula for the total torque:
\begin{equation}
\Gamma/\Gamma_0=-0.02-3.578\alpha,
\label{eqTiso}
\end{equation}
which can be seen as a non-linear equivalent of the 2D formula of \cite{tanaka02}:
\begin{equation}
\Gamma_\mathrm{lin}/\Gamma_0=-1.16-2.828\alpha.
\label{eqTisolin}
\end{equation}
For a constant surface density disc ($\alpha=0$), inward migration has slowed down by a factor of $100$ through non-linear effects. However, these formulae are of limited use due to the lack of gravitational softening. In general $b$ should be of the order of $h$ to account for 3D averaging effects, which would lead to a different value of $\xs$ \citep{horse} and a different Lindblad torque \citep{drag}. The general conclusion that non-linear corotation torques can slow down Type I migration is still valid, however.

\section{A simple adiabatic model}
\label{secModel}
In this section, we will construct a simple model describing Type I migration in terms of a linear Lindblad torque plus the non-linear horseshoe drag. A first expression for the adiabatic horseshoe drag was proposed in \cite{paardpap08}, obtained by integrating the density perturbation due to entropy conservation over the disc. In this approach, an assumption has to be made on the exact geometry of the horseshoe region. \cite{paardpap08} considered rectangular streamlines, and showed that the resulting torque is of the correct magnitude. However, in reality streamlines will not be rectangular, which can have a large impact on the torque. Here, we try to relax this assumption and take a different approach that allows us to combine the contributions of entropy and specific vorticity in a simplified way, and include the contribution of the Lindblad torque. 

\begin{figure}
\centering
\resizebox{\hsize}{!}{\includegraphics[]{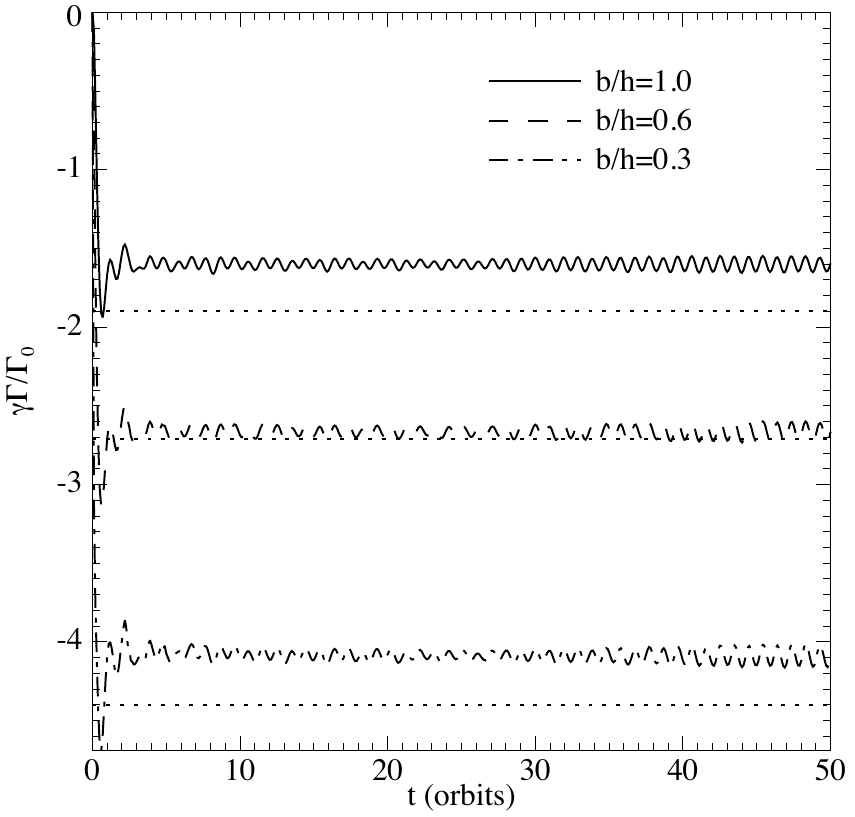}}
\caption{Total torque on a $q=1.26\cdot 10^{-5}$ planet embedded in an adiabatic disc ($\gamma=5/3$) with $\alpha=3/2$ and $\beta=1$, so that the corotation torque vanishes. Different curves denote different values of the softening parameter $b$, and the dotted lines show the prediction of equation \ref{eqLindblad}. Results were obtained with RODEO.}
\label{figlindsoft}
\end{figure}
\begin{figure}
\centering
\resizebox{\hsize}{!}{\includegraphics[]{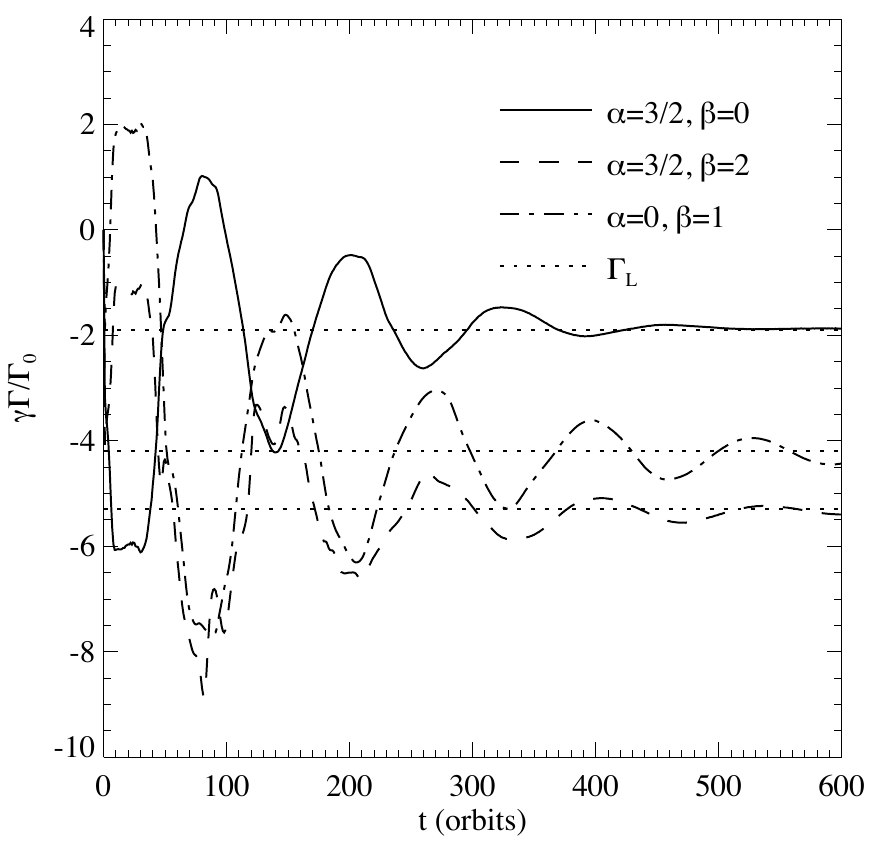}}
\caption{Total torque on a $q=1.26\cdot 10^{-5}$ planet embedded in an adiabatic disc ($\gamma=5/3$, $h=0.05$) with different density and temperature profiles, for $b/h=0.4$. Since the disc is adiabatic, the corotation torque saturates, leaving the Lindblad torque only. The dotted lines indicate the prediction of equation \ref{eqLindblad}. Results were obtained with RODEO.}
\label{figlindblad}
\end{figure}
\begin{figure*}
\begin{center}
\includegraphics[width=\textwidth]{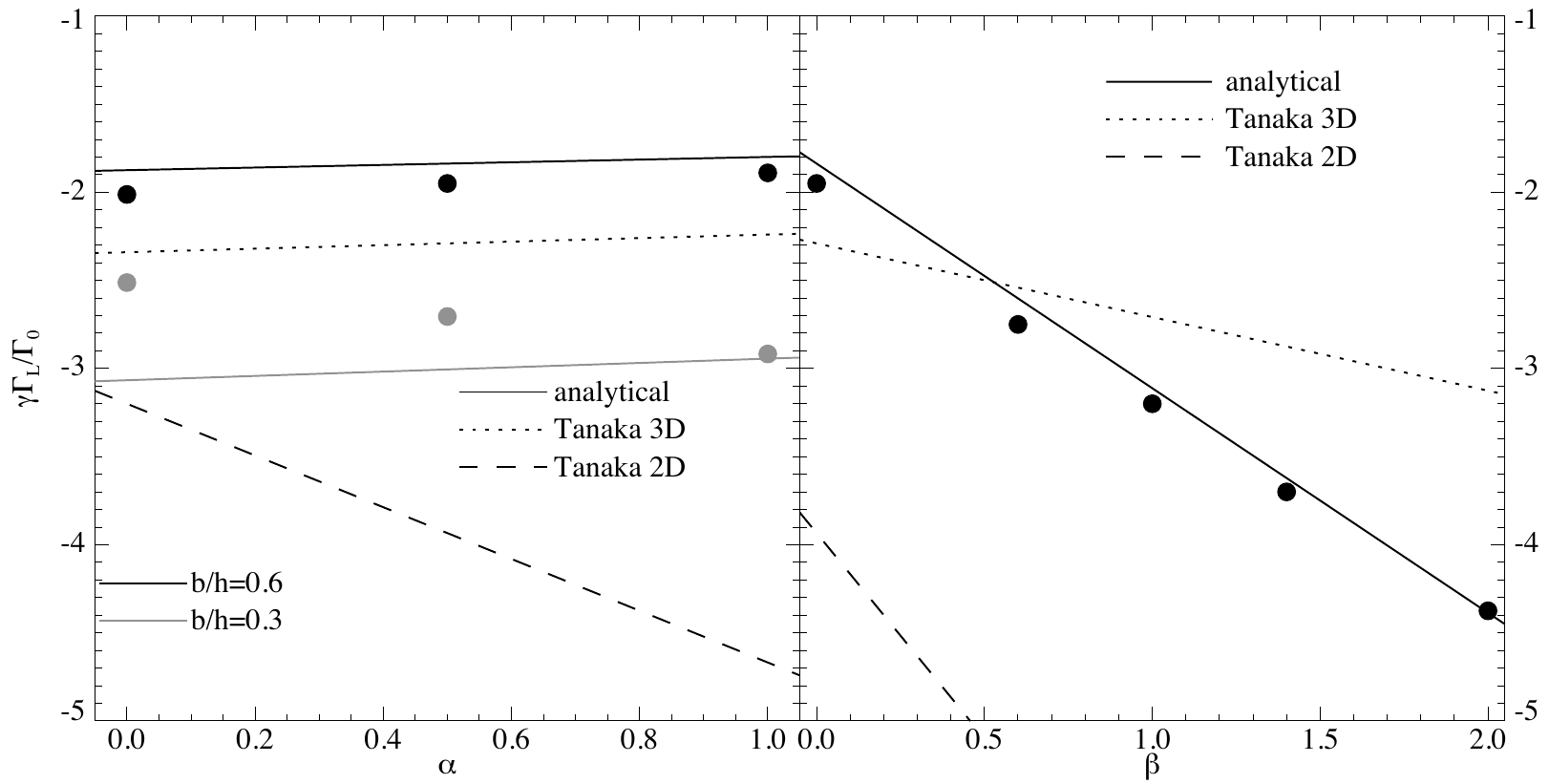}
\caption{
Lindblad torque on a $q=5\cdot 10^{-6}$ planet embedded in an adiabatic disc ($\gamma=1.4$) with $h=0.05$. Solid lines indicate the prediction of equation \ref{eqLindblad}, the dotted line indicates the 3D result of \citet{tanaka02} and the dashed line the 2D result from \citet{tanaka02}. 
Symbols denote results from numerical simulations, obtained with FARGO. Left panel: $\beta=0$, for different values of $\alpha$. Two values of $b/h$ were considered: $b/h=0.6$ (black solid line, black symbols), 
and $b/h=0.3$ (grey solid line, grey symbols). Right panel: $\alpha=1/2$ with $b/h=0.6$, for different values of $\beta$.}
\label{figlindbladalphabeta}
\end{center}
\end{figure*}

\subsection{Linear torques}
\label{secLinear}
Linearization of the two components of the equations of motion, together with the continuity equation and the adiabatic condition (equations \ref{eqcont}, \ref{eqmot} and \ref{eqent}) yields a pair of first order ordinary differential equations \citep[see][their equations 14 and 15]{paardpap08}. We have solved these linear equations for different background surface density and temperature profiles to obtain simple estimates of the Lindblad torque and linear corotation torque. 

Solving the linear equations \citep[see][]{paardpap08} results in a linear Lindblad torque:
\begin{equation}
\gamma \Gamma_\mathrm{L}/\Gamma_0=-(2.5+1.7\beta-0.1\alpha)\left(\frac{0.4}{b/h}\right)^{0.71}.
\label{eqLindblad}
\end{equation}
It was shown in \cite{paardpap08} that the linear corotation torque in adiabatic discs is associated with singularities due to radial gradients in entropy and in the quantity
\begin{equation}
\frac{\Sigma\kappa^2}{\Omega p^{2/\gamma}},
\end{equation}
with $\kappa$ the epicyclic frequency, equal to $\Omega$ in a Keplerian disc. The condition that the above quantity should be constant for the corotation torque to be zero is a generalisation of the condition that the gradient of specific vorticity should vanish, which applies in the strictly barotropic case. One is then lead to a two-term expression for the linear corotation torque; one proportional to $\xi$, and one proportional to $3/2+(1-2/\gamma)\alpha-2\beta/\gamma$. From our linear calculations, we found
\begin{eqnarray}
\gamma \Gamma_\mathrm{c,lin}/\Gamma_0=
0.7\left(\frac{3}{2}+\left(1-\frac{2}{\gamma}\right)\alpha-\frac{2\beta}{\gamma}\right)
\left(\frac{0.4}{b/h}\right)^{1.26}+ \nonumber \\
2.2\xi\left(\frac{0.4}{b/h}\right)^{0.71},
\end{eqnarray} 
which can be written in terms of $\xi$ rather than $\beta$:
\begin{eqnarray}
\gamma \Gamma_\mathrm{c,lin}/\Gamma_0=
0.7\left(\frac{3}{2}-\alpha-\frac{2\xi}{\gamma}\right)
\left(\frac{0.4}{b/h}\right)^{1.26}+ \nonumber \\
2.2\xi\left(\frac{0.4}{b/h}\right)^{0.71},
\end{eqnarray} 
making the total linear torque
\begin{equation}
\Gamma_\mathrm{lin}=\Gamma_\mathrm{L}+\Gamma_\mathrm{c,lin}.
\label{eqTlin}
\end{equation}
We will compare this linear estimate to non-linear simulations at early times in section \ref{secTest}.

For an isothermal disc ($\beta=0$, $\gamma=1$, and therefore $\xi=0$), and $b/h=0.4$, we have 
\begin{equation}
\Gamma_\mathrm{iso}/\Gamma_0=-1.4-0.6\alpha,
\end{equation}
while a 3D calculation by \cite{tanaka02} resulted in 
\begin{equation}
\Gamma_\mathrm{3D,iso}/\Gamma_0=-1.364-0.541\alpha.
\label{eqTanaka3D}
\end{equation}
Therefore, our adopted value of the smoothing length gives reasonable agreement with fully 3D calculations in the isothermal limit.

The dependence of the torque on softening can be quite complicated \citep[see][for the isothermal case]{drag}. We have chosen for a simple power law scaling that is valid around $b/h=0.4$, which is a reasonable value (see above). In Fig. \ref{figlindsoft} we show the total torque for a disc where the corotation torque vanishes, for different softening parameters. While equation \ref{eqLindblad} gives a good estimate for $b/h=0.6$ (and $b/h=0.4$, which is not shown), for more extreme values of $b/h$ the simple estimate fails. However, these extreme values are not of interest physically, since $b/h\approx0.4$ gives reasonable agreement with 3D results. For the sake of completeness, we note that the failure of equation \ref{eqLindblad} at small softening is not due to the failure of linearity, but due to the failure of the simple scaling with $b/h$ of equation \ref{eqLindblad}.

Not only does the magnitude of the Lindblad torque depend on softening, also its dependence on $\alpha$ changes. This already can be appreciated by comparing the 3D isothermal result on the Lindblad torque from \cite{tanaka02}:
\begin{equation}
\Gamma_\mathrm{L,3D}/\Gamma_0=-2.34+0.099\alpha,
\end{equation}   
to the 2D result given by equation \ref{eqTLindliniso}. The coefficient of $\alpha$ changes sign between 2D (unsoftened) and 3D calculations. We also see this change when using smaller softening parameters. Equation \ref{eqLindblad} gives good results for $b/h\approx 0.4$, however. This is illustrated in Fig. \ref{figlindblad}, where we show the long-term evolution of the total torque for inviscid, adiabatic discs with various temperature and density profiles. Since there is no viscosity or heat diffusion, the corotation torque saturates, leaving only the Lindblad torque, which is then compared to equation \ref{eqLindblad}. The agreement is very good for this value of $b/h$. Further experiments have shown that for smaller softening, the dependence on $\alpha$ is reversed, while for larger softening it is somewhat weaker. In all cases, however, the coefficient of $\alpha$ is small. 

This is illustrated in the left panel of Fig. \ref{figlindbladalphabeta}, where we compare numerical results for two different values of $b/h$ to equation \ref{eqLindblad}, as well as to the formulae of \cite{tanaka02}. For $b/h=0.6$, the coefficient of $\alpha$ is positive, and equation \ref{eqLindblad} gives good results for all values of $\alpha$. This also holds for $b/h=0.4$. For smaller values of $b/h$, the trend with $\alpha$ reverses, and equation \ref{eqLindblad} deviates from the numerical result by approximately $15\%$ for $\alpha=0$.  

In the right panel of Fig. \ref{figlindbladalphabeta}, we show the trend of the Lindblad torque with $\beta$. We have also reconstructed the $\beta$-dependence of the linear results from \cite{tanaka02} (their tables 1 and 2). Their 3D result gives a temperature dependence that is less steep than we find from our 2D simulations. The numerical results are in very good agreement with equation \ref{eqLindblad}. 

\begin{figure}
\centering
\resizebox{\hsize}{!}{\includegraphics[]{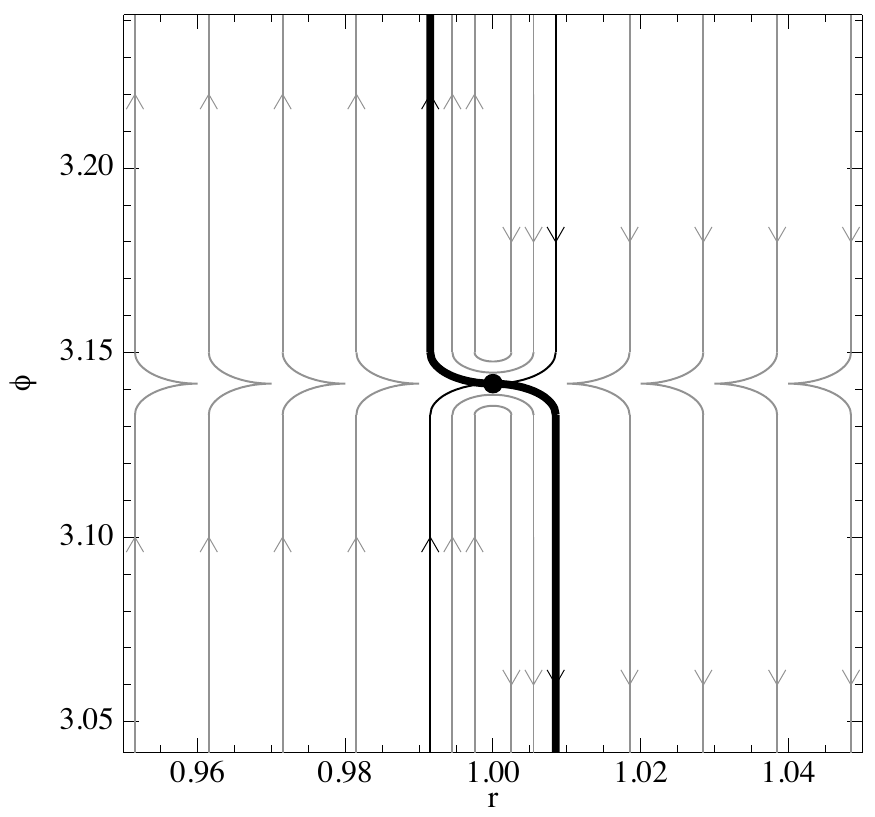}}
\caption{Schematic overview of streamlines near the horseshoe region, with the planet indicated by the black dot. The separatrix is colored black, and the thick curve indicates the location of the entropy discontinuity. The top and bottom of the figure can be thought of as $\Gamma_2$ and $\Gamma_1$, respectively. This picture applies before any material that has made the turn comes back on the other side of the planet.}
\label{fighorseschem}
\end{figure}

\subsection{Horseshoe drag}
In this section, we present a simple model for the non-linear corotation torque, the horseshoe drag, in the presence of entropy and vortensity gradients. Following \cite{ward91} we consider the torque produced by material on streamlines undergoing horseshoe turns. We consider a region ${\cal R}$  interior to the two separatrices, separating the horseshoe region from the rest of the disc, and bounded by two lines of constant $\varphi,$  $\Gamma_1$ and $\Gamma_2$ on the trailing and leading sides  of the protoplanet respectively \citep[see][and also Fig. \ref{fighorseschem}]{drag}. These boundaries are supposed to be sufficiently far from the protoplanet that the corotation torque is determined within. Assuming a steady state, this torque may be obtained by considering the conservation of angular momentum within ${\cal R}$ written in the form
\begin{equation}
\Gamma_\mathrm{c,hs}  =  \int \int_{\cal R}  
\Sigma\left( \frac{\partial \Phi_\mathrm{p}}{\partial \varphi }\right)rd\varphi dr=- \left[\int  Fdr \right]^{\Gamma_2}_{\Gamma_1},
\end{equation}
with $F=\Sigma (j - j_\mathrm{p}) (\Omega -\op) r$ \citep[see][]{drag}. Here $j=rv_{\varphi}$ is the specific angular momentum and $j_\mathrm{p}$ is $j$ evaluated at the orbital radius of the protoplanet. Assuming symmetric horseshoe turns (for a validation see section \ref{secMassCon} below), we have
\begin{equation}
\Gamma_\mathrm{c,hs}  = 2\rp \int_{0}^{\xs} (F-F_0)dx,
\label{eqHS1}
\end{equation}
where $F_0$ equals $F$ in the unperturbed disc and $x=(r-\rp)/\rp$. 
Assuming Keplerian rotation, and using first order expansions for $j-j_\mathrm{p}$ and $\Omega-\op$, we can approximate 
\begin{equation}
F-F_0\approx -\frac{3}{4}\rp^3 \sigp \op^2 x^2 \frac{\Sigma-\Sigma_0}{\Sigma_0}.
\label{eqFF0}
\end{equation}
Below, we discuss some physical arguments that allow us to relate the state (density, pressure and velocity) after the turn to the initial state. 

\subsubsection{Pressure equilibrium}
First of all, it is important to note that the disc will always try to maintain pressure balance: material that has executed a horseshoe turn should still be in pressure equilibrium with its surroundings. This assumption was not made in the original barotropic model \citep{ward91}, where the density (and therefore the pressure) was allowed to change while keeping the rotation profile fixed, similar as in equation \ref{eqFF0}. In reality, the disc will change its rotation profile in order to retain pressure balance. However, this adjustment after the turn does of course not affect the torque. We can therefore use equation \ref{eqFF0}, where only density changes are considered, to obtain the torque on the planet, but we have to keep in mind that the actual state of the fluid after the turn may well be different. Below, we work out the density changes due to entropy conservation and vortensity evolution. While entropy conservation can work directly on the density, because the pressure can be kept constant, the vortensity evolution will mainly affect the rotation profile, just as in the barotropic case. To obtain the torque, however, it is again sufficient to consider changes in surface density only, and use equation \ref{eqFF0}.  

\subsubsection{Mass conservation}
\label{secMassCon}
One important constraint that has not been discussed so far is conservation of mass, which states that the amount of mass that goes into the horseshoe turn at $x<0$ must come out of the turn at $x>0$. If material between $x=x_0<0$ and $x=0$ will make the turn and come out of the turn between $x=0$ and $x=x_1>0$, then we must have that   
\begin{equation}
\int_{x_0}^{x_1}\Sigma(\Omega-\op)rdr=0.
\end{equation}
It is clear that the assumption of pressure balance, together with mass conservation, will lead to an asymmetry in the horseshoe leg (i.e. $x_1 \neq -x_0$). Since the effect works in the opposite direction for the other horseshoe leg, the end result is that one leg will appear wider than the other. Although this in general can affect the torque, we show below that the impact on the torque is usually quite small.

For simplicity, we take the gradients of entropy and specific vorticity to be zero. There is still an asymmetry in this case due to curvature, and this will give us an estimate of the importance of this effect. We then have $\Sigma=\Sigma_0$ and $\Omega=\okep$ (ignoring the radial pressure gradient), and using a first-order Taylor expansion of $\Omega-\op$ and $\Sigma_0$, we have
\begin{equation}
\int_{x_0}^{x_1}(1-\alpha x+x)xdx=0.
\end{equation}
Writing $x_1=-x_0+\delta x$, and keeping only terms that are first order in $\delta x$, we can solve for $\delta x$: 
\begin{equation}
\delta x=\frac{2}{3}\left(\alpha-1\right)x_0^2.
\end{equation}
Since $\delta x \ll x_0$ (because $x_0 \ll 1$), the combined effect of both horseshoe legs on the torque, which scales as $\xs^4$, is a change of a factor $(1+2\delta x/x_0)$, which can be of the order of $10\%$ for considerable gradients in density. When gradients of specific vorticity and entropy are present, they can also contribute to the asymmetry (for example through equation \ref{eqSigma} below). For gradients in entropy and vortensity that are not too large, this effect is small. The assumption made in equation \ref{eqHS1} above that the horseshoe turns are symmetric should therefore give reasonable results.
 
 \subsubsection{Entropy conservation}
In pressure equilibrium, changes in entropy are directly related to changes in density. Consider material that has made a horseshoe turn from $-x$ to $x$ (we assume symmetry at this point; see section \ref{secMassCon}). The old state at $x$ is given by $\Sigma_0$, $p_0$ and $s_0$, the new state by $\Sigma$, $p$, $s$. Pressure balance dictates that $p=p_0$, while entropy conservation gives $s=s_0(1+2\xi x)$, which makes the density
$\Sigma=(p_0/s)^{1/\gamma}$, or:
\begin{equation}
\Sigma=\left\{
\begin{array}{ll}
\Sigma_0\left(1-2\frac{\xi}{\gamma}x\right) & 0<x<\xs \\
\Sigma_0 & \mathrm{otherwise.}
\end{array}
\right.
\label{eqSigma}
\end{equation}
A similar equation holds for the other horseshoe leg. 

\begin{figure}
\centering
\resizebox{\hsize}{!}{\includegraphics[]{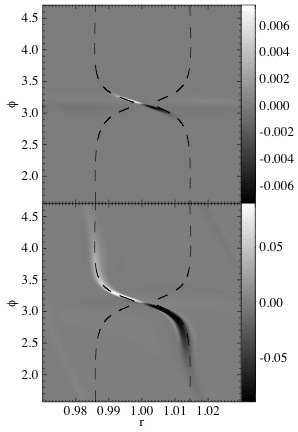}}
\caption{Top panel: source term in the vortensity equation, in units of $\op^2/\sigp$. Bottom panel: relative change in vortensity $(\omega/\Sigma-\omega_0/\Sigma_0)/(\omega_0/\Sigma_0)$. The dashed curves show the location of the separatrix. Both panels show results for a $q=1.26\cdot 10^{-5}$ planet, embedded in an $h=0.05$ disc with $\alpha=3/2$ and $\beta=-3/2$ after $10$ orbits. A large value for softening parameter was used, $b/h=2$, to reduce the influence of the wakes on the coorbital region. Results were obtained with RODEO.}
\label{figsourcevort}
\end{figure}

\subsubsection{Specific vorticity}
In a barotropic disc, specific vorticity is conserved along streamlines. We can, analogous to the entropy case discussed above, write down an expression for the change in specific vorticity after the turn:
\begin{equation}
\frac{\omega/\Sigma-\omega_0/\Sigma_0}{\omega_0/\Sigma_0}=2x\frac{d\log(\omega_0/\Sigma_0)}{d\log r},
\label{eqperturbvortbaro}
\end{equation}
for $0<x<\xs$ and zero otherwise. Here, $\omega=\nabla \times {\bf v}$ denotes the vorticity.

In a non-barotropic disc, specific vorticity, or vortensity, is no longer conserved along streamlines:
\begin{equation}
\label{eqvortp}
\frac{D}{Dt}\left(\frac{\omega}{\Sigma}\right)=\frac{\nabla \Sigma \times \nabla p}{\Sigma^3}=-\frac{\nabla s \times \nabla p}{\gamma \Sigma^2 s}.
\end{equation}
In a barotropic disc, in which $p=p(\Sigma)$, $\nabla \Sigma$ and $\nabla p$ are parallel everywhere, which makes the source term in above equation zero, with the result that vortensity is conserved along streamlines. Note, that for a non-barotropic disc, in regions where the density and pressure are smooth, the right-hand side of equation \ref{eqvortp} is small, since $p\sim h^2\Sigma$ with $h \ll 1$. However, at the outgoing separatrix, material that has made a horseshoe turn meets disc material that still has the unperturbed value of the entropy. At this specific streamline, a large entropy gradient exists (it is formally infinite across the separatrix) perpendicular to the flow (see Fig. \ref{fighorseschem}). The change in vortensity induced by the associated source term gives rise to an additional entropy-related torque. 

This is further illustrated in Fig. \ref{figsourcevort}, where we show the 2D distribution of the specific vorticity (bottom panel), together with the source term in equation \ref{eqvortp} (top panel). The background surface density profile is such that the vortensity is constant, initially. All structure seen in the bottom panel of Fig. \ref{figsourcevort} is therefore due to the source term depicted in the top panel. It is clear that this source term only acts on the outgoing separatrix, where advection of entropy generates an entropy discontinuity. The pressure gradient entering equation \ref{eqvortp} is due to the hydrostatic envelope of the planet, hence the source term is localized around the planet. Entropy advection along the horseshoe bend does not change the pressure \citep{paardpap08,baruteau08}, so the only other possible pressure gradients are due to the wakes (which play no role for low-mass planets, where $\xs < h$), or due to a global radial pressure gradient. The latter can indeed contribute to the vortensity source, but it is easy to see that the effect will be symmetric in both horseshoe legs, and therefore this does not result in a torque onto the planet. In Fig. \ref{figsourcevort}, the initial pressure was taken to be constant $(\beta=-\alpha=-3/2)$ for clarity. 

It can be seen from Fig. \ref{figsourcevort} that the lines of constant specific vorticity are not exactly parallel to the separatrix, moving slightly away from the planet's orbit near $\varphi=2.0$ and $\varphi=4.5$. This is due to the adjustment of the disc to maintain pressure equilibrium. The propagation of the vortensity discontinuity triggers a pressure wave that can be clearly identified especially at early times. It can also be observed in barotropic discs whenever there is an initial radial gradient in specific vorticity. The presence of the vortensity discontinuity makes this pressure wave more apparent.
 
From now on, we assume that vortensity is conserved everywhere except along the outgoing separatrix. Consider the integration of equation \ref{eqvortp} along the outermost streamline of the horseshoe region. The only entropy gradient of importance is perpendicular to the streamline \emph{after} it has encountered the planet (see above). We therefore write equation \ref{eqvortp} as:
\begin{equation}
\label{eqvortpot}
\frac{D}{Dt}\left(\frac{\omega}{\Sigma}\right)=-\frac{\nabla_\bot s \nabla_\parallel p}{\gamma \Sigma^2 s}=-\frac{\nabla_\bot s \nabla_\parallel \Pi}{\gamma \Sigma s},
\end{equation}
where $\nabla_\bot$ and $\nabla_\parallel$ indicate the component of the gradient perpendicular and parallel to the streamline, respectively, and $\Pi$ is the fluid enthalpy. The orientation is such that we take the gradient of $\Pi$ in the direction away from the planet, which means that we have to take the gradient of $s$ \emph{into} the horseshoe region.

We take the gradient of $s$ into the horseshoe region to be infinite, formally:
\begin{equation}
\nabla_\bot s/s=2\xi\xs\delta(x-\xs).
\end{equation}
It is easy to see that this models a jump in entropy across the separatrix that has the correct magnitude.

Noting that  factors involving $s$ are constant along the streamline, and assuming $\Sigma\approx\sigp$ and taking the velocity along the streamline to be $v=\bar v \rp\op\xs$, with $\bar v$ a constant, we can integrate equation \ref{eqvortpot} from the turn at $x=0$ near the stagnation point to a point far away from the planet to end up with:
\begin{equation}
\Delta\left(\frac{\omega}{\Sigma}\right)(x)=\frac{2\xi}{\bar v\gamma}\frac{\Pi_\mathrm{turn}-\Pi_0(\xs)}{\rp^2\op\sigp}\delta(x-\xs),
\label{eqVort1}
\end{equation}
where $\Pi_\mathrm{turn}$ is the enthalpy at the location of the turn, near the stagnation point. Along this streamline, the velocity varies from $0$ at the stagnation point to $v=3\xs\rp\op/2$ far away from the planet; therefore we need $0<\bar v < 3/2$. The choice of $\bar v$ basically depends on the exact geometry of the horseshoe region (see below). Equation \ref{eqVort1} then gives the the vortensity production at the outgoing separatrices, a process that does not operate in barotropic discs.

Note that in the case of multiple stagnation points close to the planet \citep[][see also Fig. \ref{figstream}]{masset06,horse}, there is always a single stagnation point where the entropy discontinuity starts for both horseshoe legs (the stagnation points below the planet in the top panels of Fig. \ref{figstream}). We therefore do not have to make any additional assumptions on the detailed flow topology close to the planet. 

To make further progress, we now assume that the pressure (or enthalpy) structure is not significantly changed from the barotropic case, or, equivalently, the case with $\xi=0$. For $\xi \neq 0$, advection of entropy leads to changes in enthalpy, which is then discontinuous across the separatrix. It is difficult to see which value of $\Pi$ to take (inside or outside the separatrix) in that case. Note, however, that the jump in $\Pi$ is of order $\xs$ and therefore small. It should not affect the pressure structure near the outgoing separatrix. This has been verified using numerical simulations. We can then find $\Pi_\mathrm{turn}-\Pi_0$ by using the Bernoulli invariant $E$ \citep[see][]{horse}:
\begin{equation}
E=\frac{1}{2}\rp^2(\Omega-\op)^2 + \Pi + \Phi_\mathrm{p}-\frac{3}{2}\rp^2\op^2 x^2,
\end{equation}
Considering two points on the same streamline, one at the turn (where $x=0$ and $\Omega=\op$) and one far away from the planet (where $x=\xs$ and $\Phi_\mathrm{p}=0$), we have:
\begin{equation}
\Pi_\mathrm{turn}+\Phi_\mathrm{p,turn}=-\frac{3}{8}\xs^2\rp^2\op^2+\Pi_0(\xs).
\label{eqPistag}
\end{equation}
Using equation \ref{eqPistag} in equation \ref{eqVort1}, we end up with
\begin{equation}
\Delta\left(\frac{\omega}{\Sigma}\right)(x)=\frac{2\xi}{\bar v\gamma}\frac{\op}{\sigp}\left(\frac{q}{d}-\frac{3}{8}\xs^2\right)\delta(x-\xs),
\end{equation}
with $d=\sqrt{|{\bf r}_\mathrm{turn}-{\bf r}_\mathrm{p}|^2/\rp^2+b^2}$.
Numerical simulations for isothermal discs \citep{masset06,horse} show that the stagnation point is located roughly $3/2$ softening lengths away from the planet, making $d=\sqrt{13/4}b$. In adiabatic discs, the situation is slightly different. As was shown in \cite{masset06}, the width of the horseshoe region is directly related to the perturbed value of the Bernoulli invariant at the stagnation point $E'=\Pi'_\mathrm{turn}+\Phi_\mathrm{p,turn}$:
\begin{equation}
\xs=\frac{1}{\rp\op}\sqrt{-\frac{8}{3}E'}.
\end{equation}
On the other hand, it was shown in \cite{horse} that $\xs^4 \propto 1/\gamma$. Therefore, there should exist a direct relationship between the location of the stagnation point and $\gamma$. We should have $E' \propto 1/\sqrt{\gamma}$, and, assuming for simplicity that both the perturbed enthalpy and the planet potential at the stagnation point have the same dependence on $\gamma$, this leads to
\begin{equation}
\frac{|{\bf r}-{\bf r}_\mathrm{turn}|}{\rp}=\sqrt{n\gamma-1}b,
\end{equation}
for some constant $n$. It was noted in \cite{horse} that the location of the stagnation point is essentially determined by the Lindblad wakes, which makes it very difficult to model. The best we can do is fix the constant $n$ so that we recover the isothermal result for $\gamma=1$. This means we take $n=13/4$, and we will see in section \ref{secstream} that this gives reasonably good results for $\gamma>1$. We then have 
\begin{equation}
d=\sqrt{\frac{13\gamma}{4}}b.
\label{eqd}
\end{equation}

Away from the outgoing separatrices, the specific vorticity source term is small, and we can assume conservation of vortensity. We then have for the total vortensity:
\begin{equation}
\frac{\omega}{\Sigma}=\left\{
\begin{array}{ll}
\frac{\omega_0}{\Sigma_0}\left(1-2\frac{d\log \left(\frac{\omega}{\Sigma}\right)}{d\log r}x\right) + \Delta\left(\frac{\omega}{\Sigma}\right) & 0 < x<\xs, \\
\frac{\omega_0}{\Sigma_0} & \mathrm{otherwise.}
\end{array}
\right.
\label{eqperturbvort}
\end{equation}
The term in parenthesis results from conservation of specific vorticity, that is also present in barotropic discs.   
 
\subsubsection{Total horseshoe drag}
We now add the contributions of entropy (equation \ref{eqSigma}) and vortensity (equation \ref{eqperturbvort}) to the density perturbation, which gives
\begin{equation}
\frac{\Sigma-\Sigma_0}{\Sigma_0}=-2\frac{\xi}{\gamma}x+2\left(\alpha-\frac{3}{2}\right)x-\frac{\Sigma_0}{\omega_0}\Delta\left(\frac{\omega}{\Sigma}\right).
\label{eqDenstot}
\end{equation}
The vortensity-related perturbation (the second term on the right-hand side) is the same as in the barotropic case. The entropy-related density perturbation (the first and the last term on the right-hand side) is caused by density structures produced by material conserving its entropy, bound to the horseshoe region, plus an additional component linked to the production of vortensity at the outgoing separatrices. We stress again that the contribution of the vortensity to the density perturbations (the last two terms in the equation above) will affect the rotation profile rather than the density. However, the torque exerted on the planet will be the same, just as in the barotropic case.

In order to find the torque, we can now perform the integral in equation \ref{eqHS1}, using equations \ref{eqFF0} and \ref{eqDenstot}, yielding
\begin{equation}
\Gamma_\mathrm{c,hs}=\frac{3}{4}\sigp\rp^4\op^2\xs^4\left(\frac{3}{2}-\alpha+\frac{\xi}{\gamma}\left(\frac{8q}{\bar v d\xs^2}+1-\frac{3}{\bar v}\right)\right)
\end{equation}
For $\xi=0$, we recover the barotropic result of \cite{ward91}. Note that although $\xs$ is well-defined for $b\rightarrow 0$ \citep{horse}, the contribution of the entropy discontinuity diverges if $d \rightarrow 0$ as well. 

If we write $\xs=C\sqrt{q/h}/\gamma^{1/4}$, with $C=C(b/h)$ \citep{horse}, then we have:
\begin{equation}
\gamma\Gamma_\mathrm{c,hs}/\Gamma_0=\frac{3}{4}C^4\left(\frac{3}{2}-\alpha+\frac{\xi}{\gamma}\left(\frac{8\sqrt{\gamma}}{\bar v C^2}\frac{h}{d}+1-\frac{3}{\bar v}\right)\right).
\label{eqHSadi}
\end{equation}
Note that the component due to a radial entropy gradient can easily overpower the contribution from the vortensity gradient. Note also that for fixed $b/h$ (which makes $C$ a constant as well, as long as $\xs<h$), the horseshoe drag $\Gamma_\mathrm{c,hs}$ scales as $q^2/h^2$, just as the linear torque.   

Numerical simulations and analytical arguments indicate that the geometry of the horseshoe region is the same for all low-mass planets, as long as $\xs < h$ \citep{masset06,horse}. Therefore, one choice of $\bar v$ should suffice. We have obtained good agreement with numerical simulations using  $\bar v =1.0$. We have measured the horseshoe width $\xs$ to be
\begin{equation}
\xs=\frac{1.1}{\gamma^{1/4}}\left(\frac{0.4}{b/h}\right)^{1/4}\sqrt{\frac{q}{h}},
\label{eqxs}
\end{equation}
i.e. $C=1.1$ for $b=0.4 h$. We note that the scaling with $b/h$ breaks down for small softening ($b/h<0.3$). Equation \ref{eqHSadi} is then completely determined:
\begin{eqnarray}
\gamma\Gamma_\mathrm{c,hs}/\Gamma_0=1.1\frac{0.4}{b/h}\left(\frac{3}{2}-\alpha\right)+\frac{\xi}{\gamma}\frac{0.4}{b/h}\left(10.1\sqrt{\frac{0.4}{b/h}} -2.2\right).
\label{eqHSadi2}
\end{eqnarray}

\begin{figure}
\centering
\resizebox{\hsize}{!}{\includegraphics[]{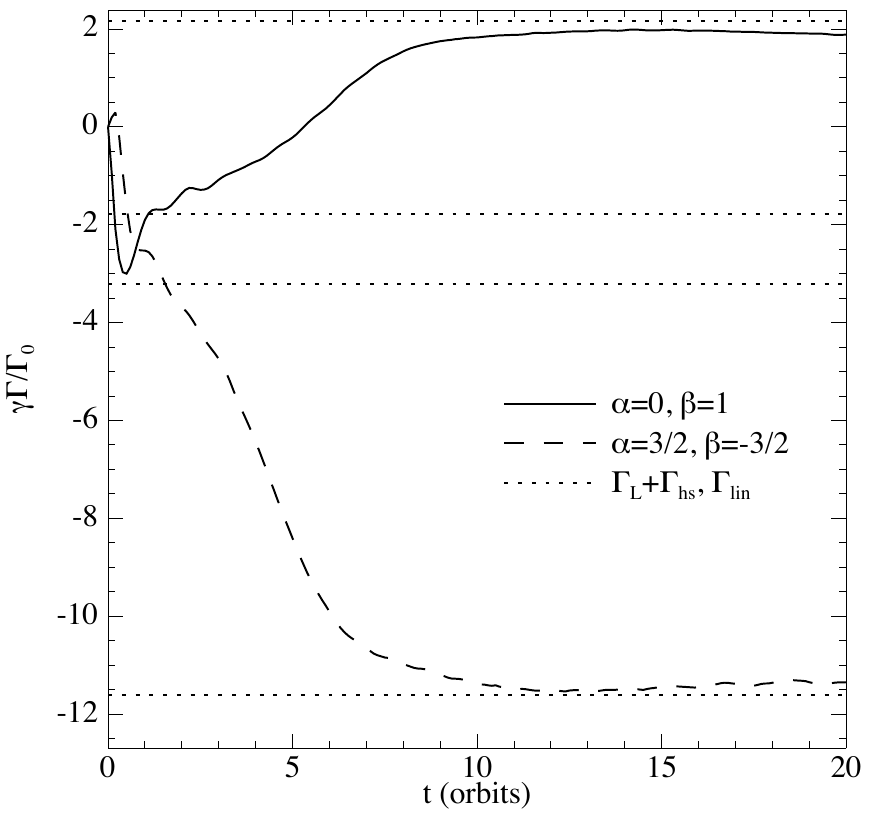}}
\caption{Total torque on a $q=1.26\cdot 10^{-5}$ planet (with $b/h=0.4$) embedded in an adiabatic disc ($\gamma=5/3$, $h=0.05$) with different density and temperature profiles. The dotted lines indicate the prediction of equation \ref{eqTlin} (middle two lines) and the result of equation\ref{eqTtot} (top and bottom lines). Results were obtained using RODEO.}
\label{fignonlincomp}
\end{figure}

\begin{figure}
\centering
\resizebox{\hsize}{!}{\includegraphics[]{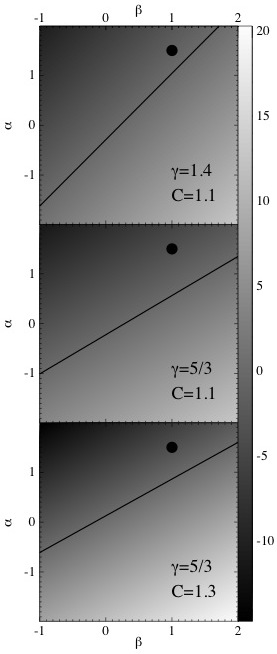}}
\caption{Total torque, as given by equation \ref{eqTtot}, in units of $\Gamma_0/\gamma$. The solid line indicates where the total torque is zero. The black dot indicates the Minimum Mass Solar Nebula, having $\beta=1$ and $\alpha=3/2$. Top panel: $\gamma=1.4$, middle panel: $\gamma=5/3$. The bottom panel is the same as the middle panel, but for $C=1.3$, valid for higher mass planets that obtain the maximum value of $\xs$.}
\label{figtorquetot}
\end{figure}

\subsection{Total torque}
\label{secTotaltorque}
The total torque in the non-linear regime ($t>\sim 2$ orbits in an inviscid disc), before saturation sets in, is given by \citep{drag}
\begin{equation}
\Gamma=\Gamma_\mathrm{L}+\Gamma_\mathrm{c,hs},
\label{eqTtotgen}
\end{equation}
with $\Gamma_\mathrm{L}$ given by equation \ref{eqLindblad}, and $\Gamma_\mathrm{c,hs}$ given by equation \ref{eqHSadi2}.

We now take $b/h=0.4$, which makes the total torque:
\begin{eqnarray}
\gamma\Gamma/\Gamma_0=-2.5-1.7\beta+0.1\alpha+
1.1\left(\frac{3}{2}-\alpha\right)+7.9\frac{\xi}{\gamma},
\label{eqTtot}
\end{eqnarray}
where the last two terms describe the non-linear corotation torque. This equation is compared to numerical simulations in Fig. \ref{fignonlincomp}, showing remarkably good agreement between our simple model and fully non-linear hydrodynamical simulations. We will defer a detailed numerical analysis to Sect. \ref{secTest}.

Equation \ref{eqTtot} is shown as a function of $\alpha$ and $\beta$ in the middle panel of Fig. \ref{figtorquetot} for $\gamma=5/3$. Positive torques, and therefore outward migration, are readily obtained for $\beta>0$, i.e. for temperature profiles that decrease outward. A glance at equation \ref{eqLindblad} reveals that this is completely due to the entropy-related corotation torque, since the Lindblad torque becomes more negative with increasing $\beta$. In the top panel of Fig. \ref{figtorquetot} we show equation \ref{eqTtot} for $\gamma=1.4$, a value often adopted for protoplanetary discs \citep{paardpap08,kley08}. For this lower value of $\gamma$, the entropy gradient depends stronger on the temperature gradient (see equation \ref{eqxi}), resulting in a steeper slope for the zero-torque line. Since protoplanetary discs are expected to have $\beta>0$ in most parts, this indicates that outward migration is a serious possibility.

In the bottom panel of Fig. \ref{figtorquetot}, we return to $\gamma=5/3$, but use a larger value of $C$, $C=1.3$, that would correspond to higher mass planets that are able to push against the Lindblad wake to take the stagnation point close to $(r,\varphi)=(\rp,\pi)$ \citep{masset06}. Formally, one should use $d=b$ as well, but we have found that the stagnation point never actually reaches the planet. Effectively, one should use $C<1.3$ in combination with $b<d<\sqrt{5}b$, but $C=1.3$ in combination with $d=\sqrt{5}b$ gives reasonably good results (see Sect. \ref{secHighmass}).  

\begin{figure}
\centering
\resizebox{\hsize}{!}{\includegraphics[]{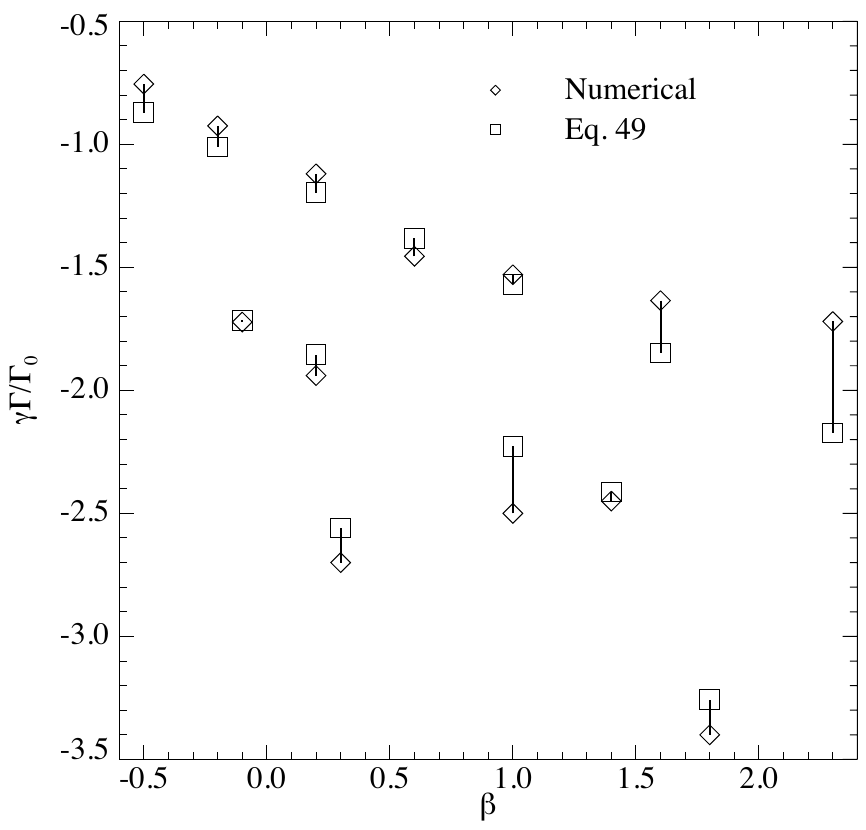}}
\caption{Total torque on a $q=5\cdot 10^{-6}$ planet (with $b/h=0.6$) embedded in a locally isothermal disc with $h=0.05$. Diamonds denote numerical results obtained with FARGO, squares the result of equation \ref{eqlociso}. Corresponding results are connected with a line. For the top row (7 runs), $\alpha=1/2$, for the middle row (4 runs) $\alpha=3/2$, and for the bottom row (2 runs) $\alpha=5/2$.}
\label{figformulaiso}
\end{figure}

\begin{figure}
\centering
\resizebox{\hsize}{!}{\includegraphics[]{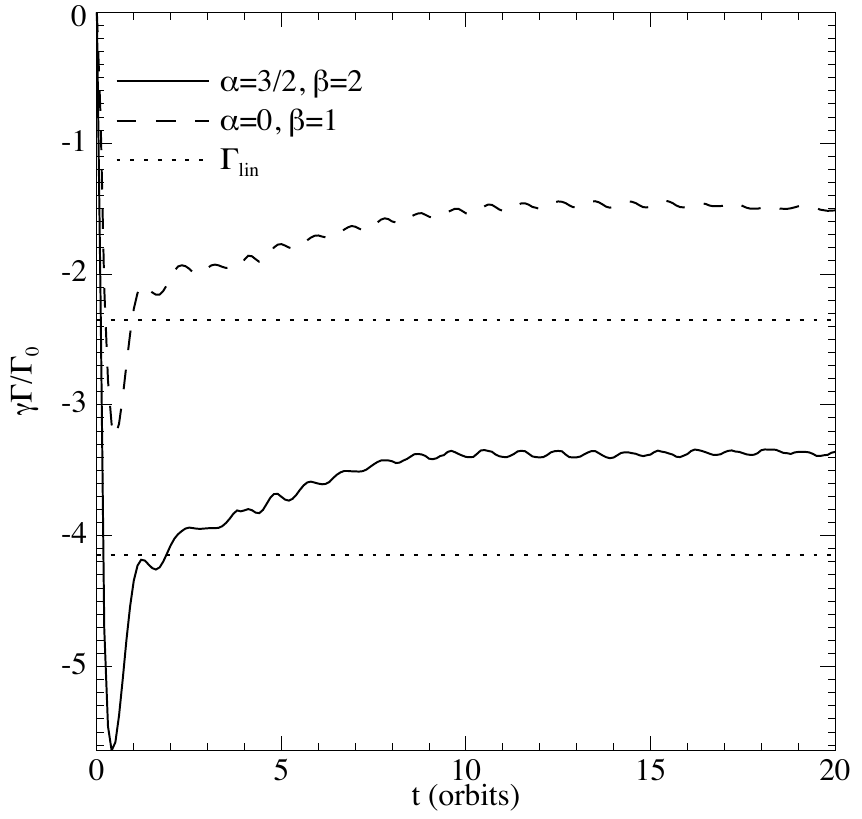}}
\caption{Total torque on a $q=1.26\cdot 10^{-5}$ planet (with $b/h=0.4$) embedded in a locally isothermal disc ($\gamma=1$, $h=0.05$, $\beta \neq 0$) with different density and temperature profiles. The dotted lines indicate the result of equation \ref{eqTlin}. Results were obtained with RODEO}
\label{figlociso}
\end{figure}

\subsection{Locally isothermal limit}
It is straightforward to obtain the isothermal limit of equation \ref{eqTtot} by setting $\beta=0$ and $\gamma=1$ (and therefore $\xi=0$), which then leaves the linear Lindblad torque plus the vortensity-related horseshoe drag:
\begin{equation}
\Gamma_\mathrm{iso}/\Gamma_0=-0.85-\alpha,
\end{equation}
where we have used $b/h=0.4$. Compared to the 3D linear result (equation \ref{eqTanaka3D}), migration has slowed down by approximately $50\%$ for $\alpha=0$, but the direction is inward for all realistic surface density profiles.

A different approximation that is often used is the \emph{locally} isothermal limit, which means solving the isothermal equations but with a radially varying sound speed (or, equivalently, temperature). One arrives at this limit by taking $\gamma \rightarrow 1$, and, crucially, invoke infinitely efficient thermal diffusion. This effectively takes the entropy-related horseshoe drag into the linear regime. The total torque then consists of the linear Lindblad torque, the linear entropy-related corotation torque plus the non-linear vortensity-related horseshoe drag:
\begin{eqnarray}
\Gamma_\mathrm{lociso}/\Gamma_0=-(2.5-0.5\beta-0.1\alpha)\left(\frac{0.4}{b/h}\right)^{0.71}-\nonumber\\
1.4\beta\left(\frac{0.4}{b/h}\right)^{1.26}+1.1\left(\frac{3}{2}-\alpha\right)\left(\frac{0.4}{b/h}\right).
\label{eqlociso}
\end{eqnarray}
Note that the temperature dependences of the Lindblad and corotation torque work against each other, leaving a total torque that is less sensitive to temperature variations.  

In Fig. \ref{figformulaiso}, we compare equation \ref{eqlociso} to numerical results obtained with FARGO, showing good agreement over a wide range of $\beta$. Note that, contrary to the adiabatic case, a negative temperature gradient works in favour of inward migration. This is due to the relatively weak dependence of the linear corotation torque on $\beta$, compared to the Lindblad torque. We have found inward migration for all reasonable values of $\alpha$ and $\beta$.

However, the good agreement as seen in Fig. \ref{figformulaiso} is not the whole story, as can be seen from Fig. \ref{figlociso}. While the solid curve denotes a case of constant specific vorticity, a non-linear rise in the torque can be observed. This is due to the source term in the vorticity equation. It is important to note that the analysis presented above for adiabatic discs is not valid in the locally isothermal case, since entropy (which would correspond to $\cs^2$ in this case) is not conserved along a streamline. One would get a source term proportional to $\partial p/\partial \varphi ~d\cs^2/dr$, the effect of which will strongly depend on the geometry of the horseshoe region. Note, however, that the impact of the source term is quite small even for a steep temperature gradient $\beta=2$. We note that the effect is negligible for $|\beta| < 1$, and for $\beta=2$ comparable to the non-linear effect associated with barotropic horseshoe drag (see the dashed curve in Fig. \ref{figlociso}). A locally isothermal disc is nevertheless an interesting case, since it would correspond to a part of the disc that can cool very efficiently, i.e. the outer parts of protoplanetary discs.
  
\section{Numerical results}
\label{secTest}
In this section, we will try to validate the simple model by means of numerical simulations. 

\begin{figure}
\centering
\resizebox{\hsize}{!}{\includegraphics[]{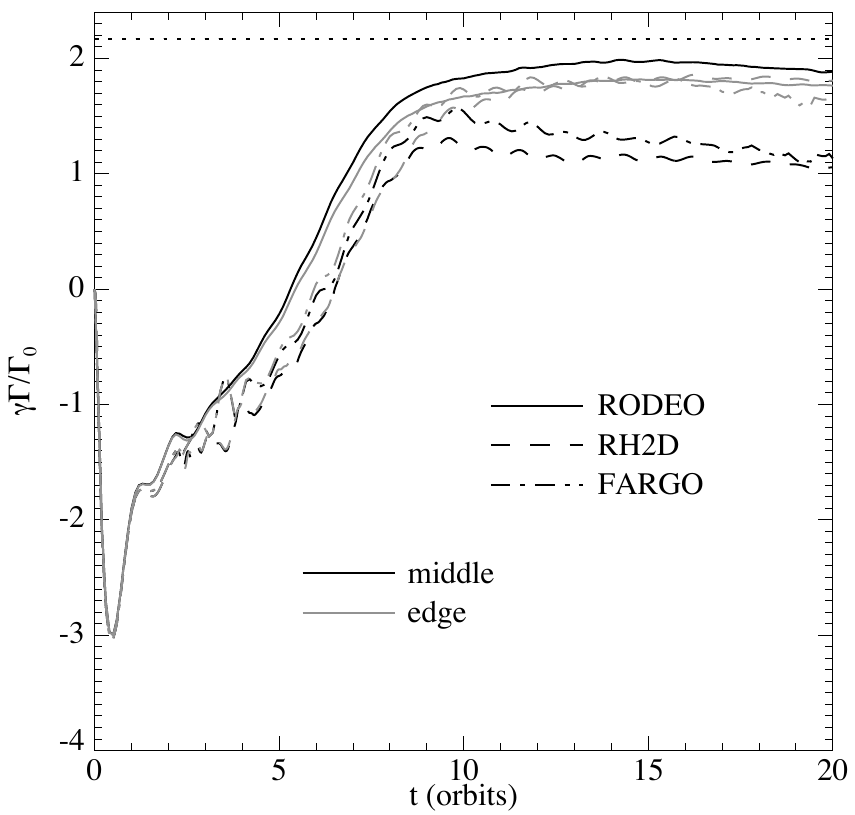}}
\caption{Total torque on a $q=1.26\cdot 10^{-5}$ planet (with $b/h=0.4$) embedded in an adiabatic disc with $\gamma=5/3$, $h=0.05$, $\alpha=0$ and $\beta=1$. Different line styles denote different codes, while different line colors indicate whether the planet is kept in the middle of a grid cell or on the edge. The dotted line indicates the result of equation \ref{eqTtot}. The resolution used is $\Delta r/\rp=\Delta \varphi=0.0025$.}
\label{figcompcodes}
\end{figure}

\begin{figure}
\centering
\resizebox{\hsize}{!}{\includegraphics[]{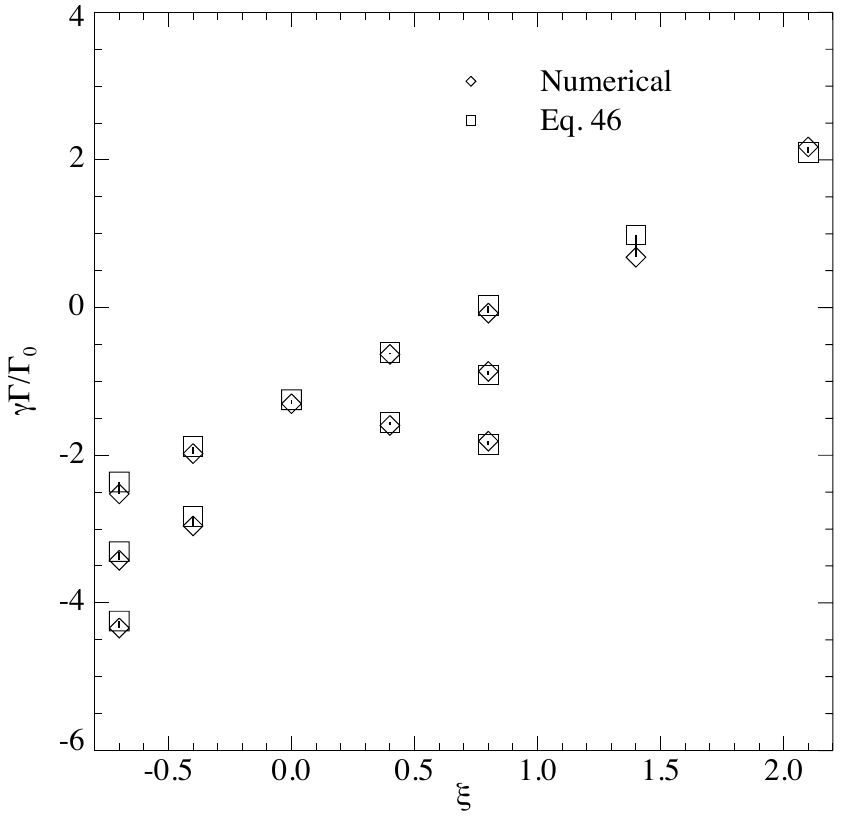}}
\caption{Total torque on a $q=5\cdot 10^{-6}$ planet (with $b/h=0.6$) embedded in an adiabatic disc with $h=0.05$ and $\gamma=1.4$. Diamonds denote numerical results obtained with FARGO, squares the result of equation \ref{eqTtotgen}. Corresponding results are connected with a line. For $\xi=-0.7$ and $\xi=0.8$, three values of $\alpha$ were considered, $\alpha=\{0.5,1.5,2.5\}$, from top to bottom. For $\xi=-0.4$ and $\xi=0.4$, we show results for $\alpha=0.5$ and $1.5$ (from top to bottom), while for all other values of $\xi$ results are shown for $\alpha=0.5$ only. In all cases the temperature gradient can be found from $\beta=\xi+0.4\alpha$.}
\label{figformulaadi}
\end{figure}

\subsection{Code comparison}
We first check whether our independent numerical methods give similar results. In Fig. \ref{figcompcodes}, we compare results for the three different numerical methods on an adiabatic disc with $\gamma=5/3$, $h=0.05$, $\alpha=0$ and $\beta=1$. The grey curves denote models for which the grid is chosen so that $r/\rp=1$ lies on the edge of a grid cell. For these models, all codes agree nicely on the final torque, with each other as well as with equation \ref{eqTtot}. They also agree on the linear part of the torque (for $t<2$). RODEO shows a slightly faster rise in the torque. Since this difference arises as soon as non-linear effects set in, it probably originates close to the separatrix, where the vorticity source term plays a major role. 

In Fig. \ref{figformulaadi}, we compare the difference between equation \ref{eqTtotgen} and numerical results obtained with FARGO for different entropy gradients. The relative difference is well within $20\%$ except for the cases where $\xi=0.8$. Results obtained with RODEO show similar good agreement. Therefore, not only do all codes agree on the torque, they also agree very well with our simple analytic model over a large range of $\alpha$ and $\beta$ within $30\%$.

Results displayed in Fig. \ref{figcompcodes} were obtained at a resolution of $\Delta r/\rp = \Delta \varphi=0.0025$. For $C=1.1$, we have $\xs=0.015$, so the half-width of the horseshoe region is resolved by 6 grid cells. Lowering this to 4 cells had little effect on the torques, suggesting that they are converged at this resolution. There is however the interesting difference between the black and grey curves in Fig. \ref{figcompcodes}, which we discuss next. 

\begin{figure}
\centering
\resizebox{\hsize}{!}{\includegraphics[]{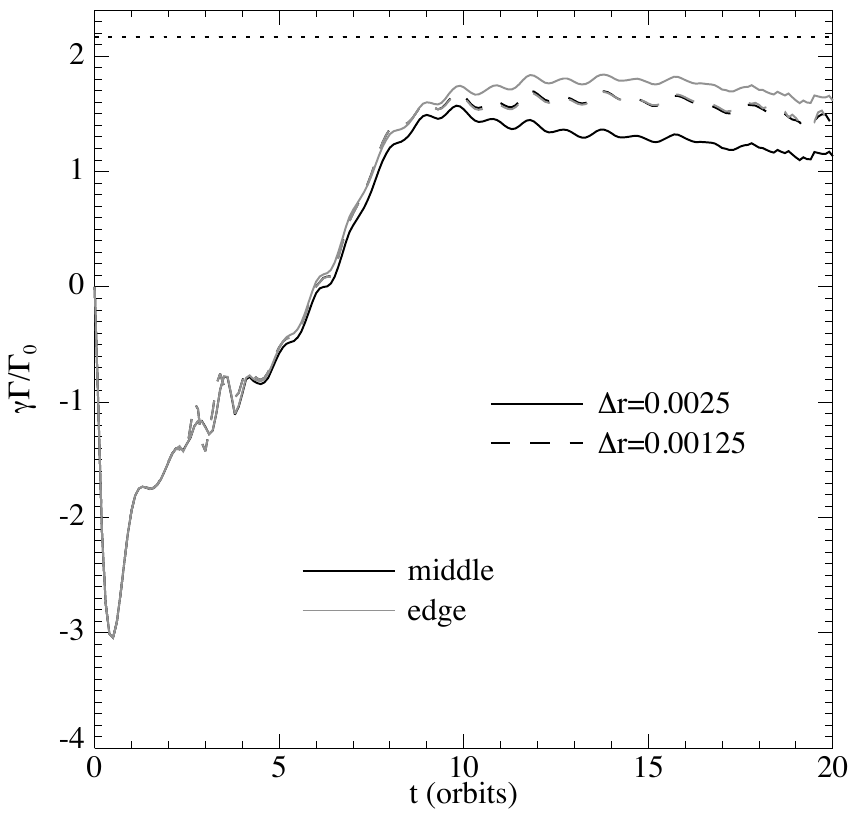}}
\caption{Same as Fig. \ref{figcompcodes}, for different radial resolutions. All results were obtained with FARGO.}
\label{figconverge}
\end{figure}

\begin{figure*}
\centering
\includegraphics[width=\linewidth]{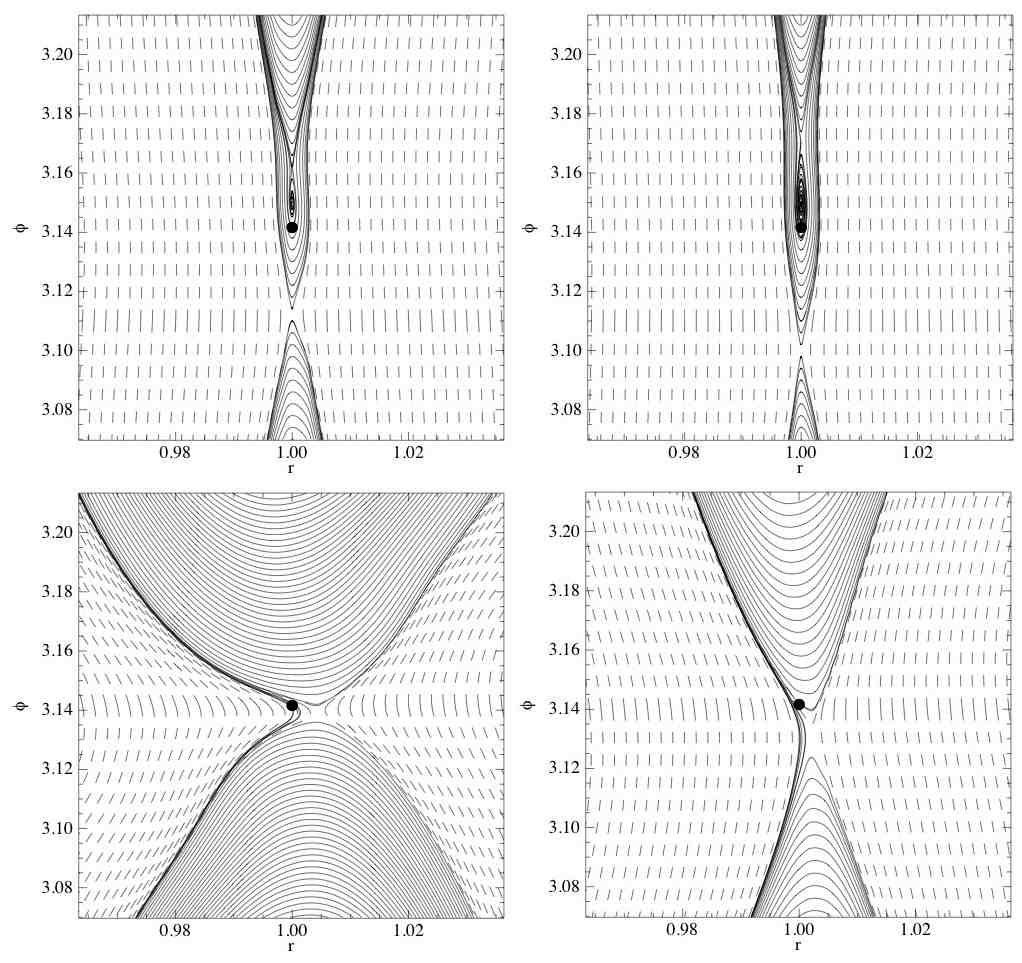}
\caption{Streamlines close to the planet embedded in a disc with $\alpha=0$ and $\beta=0$, for $\gamma=1$ (isothermal models, left panels) and $\gamma=5/3$, right panels. Top panels: $q=1.26\cdot 10^{-5}$, bottom panels: $q=1.08\cdot 10^{-4}$. Solid curves denote streamlines belonging to the horseshoe region, while dashed curves denote streamlines from the inner and outer disc. Results were obtained with RODEO.}
\label{figstream}
\end{figure*}
 
\subsection{Planet position}
The only way in which the simulations indicated by the grey and the black curves in Fig. \ref{figcompcodes} differ is the position of the planet on the grid. While for the results obtained with RODEO, this has almost no effect on the torque, both FARGO and RH2D show a difference of approximately $50\%$ between the two planet positions. This is a numerical effect that has to do with the sharp gradient in specific vorticity that arises at the outgoing separatrix under influence of the entropy-related source term. Such a sharp feature is difficult to handle numerically. It also appears in locally isothermal simulations, but is absent in fully isothermal runs, where there is no source term for the specific vorticity. The use of a Riemann solver, which is specifically designed to handle discontinuities, largely evades this problem. Being a numerical artefact, it also disappears at higher resolution, which is illustrated in Fig. \ref{figconverge}. After doubling the radial resolution, the grey curve becomes indistinguishable from the black curve. The torque then does not show the slow decay after 10 orbits, and settles close to the value that is obtained by putting the planet on the edge of a grid cell at low resolution.  

The amplitude of this numerical effect strongly depends on the detailed geometry of the horseshoe region close to the stagnation point, which in turn depends on background gradients of temperature and density. We observed that it is virtually absent for a disc with $\alpha=3/2$ and $\beta=-3/2$. 

 \subsection{Streamline analysis}
 \label{secstream}
The exact geometry of the horseshoe region is of significant importance in determining the corotation torque. In isothermal discs through $\xs$ only, but for adiabatic discs also through the location of the stagnation point ${\bf r}_\mathrm{stag}$. There is a direct relation between $\xs$ and ${\bf r}_\mathrm{stag}$ \citep{masset06,horse}. In Fig. \ref{figstream}, we show the streamlines close to the planet for isothermal models (left panels) and adiabatic models (right panels) for a $q=1.25\cdot 10^{-5}$ planet 
(top panels) and a $q=1.08\cdot 10^{-4}$ planet (bottom panels). The top left panel is consistent with the findings in \cite{masset06}, with the stagnation point located approximately $3b\rp/2$ from the planet. For an adiabatic model with the same disc parameters, the stagnation point shifts further away from the planet, to approximately $2b\rp$. It was checked that an adiabatic model with $\gamma=1.01$ looks similar to the isothermal model. We conclude that the simple scaling of equation \ref{eqd} gives a reasonable estimate for the location of the stagnation point for different values of $\gamma$. We note again that it is very difficult to model the position of the stagnation point in a simple way, because it depends on the way the wake is able to influence the corotation region \citep{horse}.

\begin{figure}
\centering
\resizebox{\hsize}{!}{\includegraphics[]{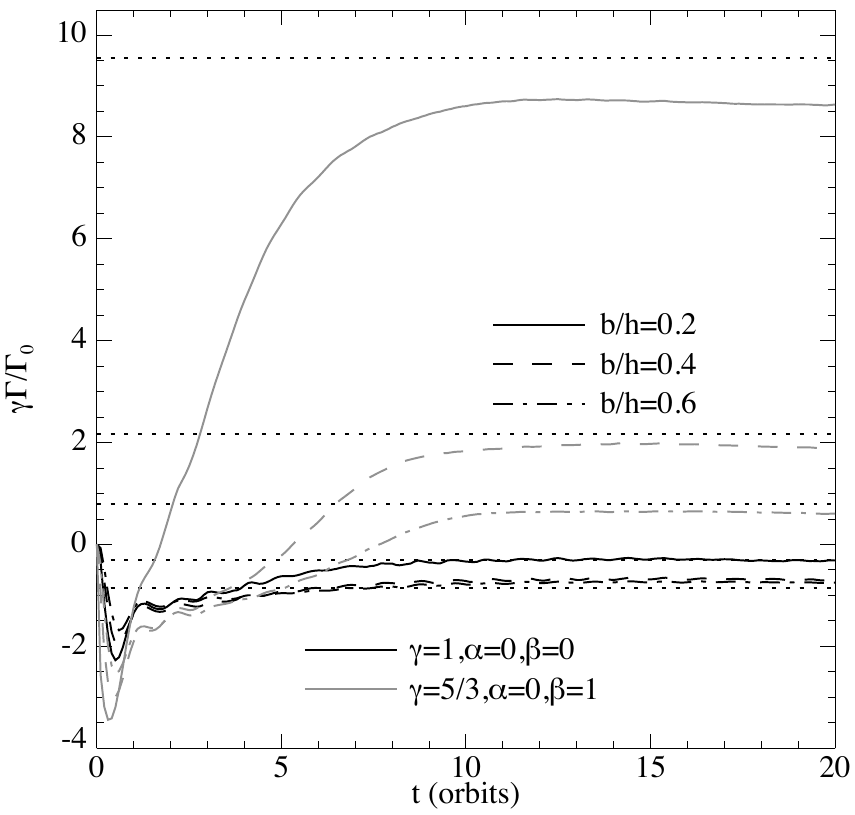}}
\caption{Total torque on a $q=1.26\cdot 10^{-5}$ planet embedded in a $h=0.05$ disc with $\alpha=0$, for different values of $b/h$, for $\gamma=1$ and $\beta=0$ (black curves), and for $\gamma=5/3$ and $\beta=1$ (grey curves). Dotted lines indicate the result of equation \ref{eqTtot}, for appropriate values of $C$ and $d$. Results were obtained with RODEO.}
\label{figsoft}
\end{figure}
\begin{figure}
\centering
\resizebox{\hsize}{!}{\includegraphics[]{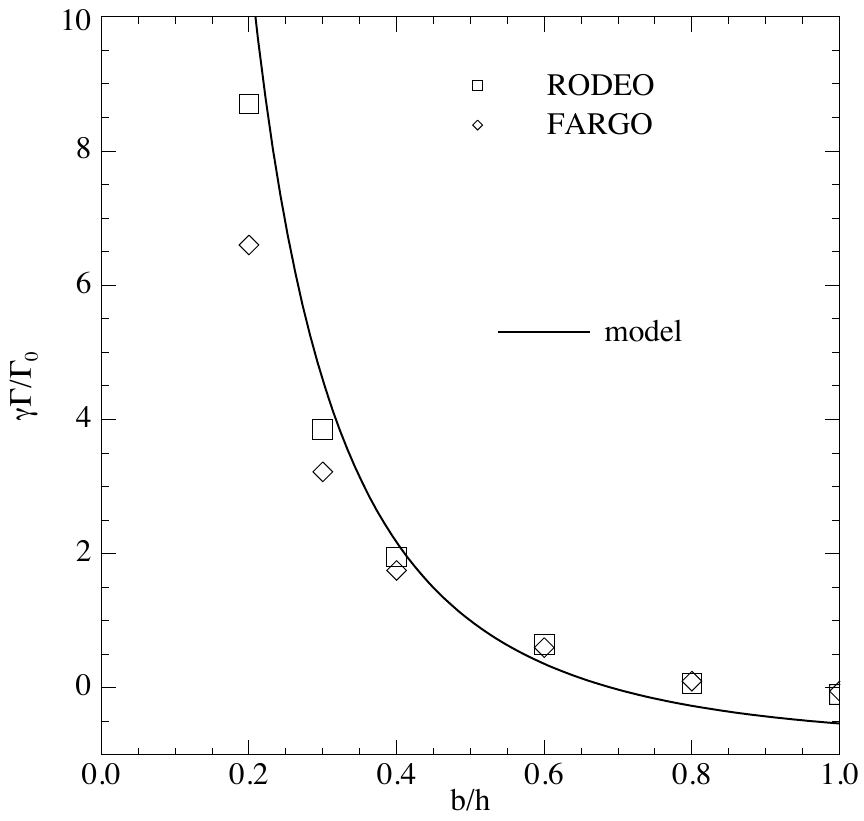}}
\caption{Total torque on a $q=1.26\cdot 10^{-5}$ planet embedded in a $h=0.05$ disc ($\gamma=5/3$) with $\alpha=0$, $\beta=1$, for different values of $b/h$. Results are shown for RODEO (squares) and FARGO (diamonds). The result of equation \ref{eqTtotgen} is indicated by the solid curve.}
\label{figsoftcomp}
\end{figure}
  
\subsection{Softening}
In this section we discuss models that use different gravitational softening parameters. Although we have argued in Sect. \ref{secLinear} that using $b/h=0.4$ gives linear torques that are in agreement with 3D linear theory, in is not clear at present what value of $b/h$ will reproduce the 3D \emph{non} linear torque (or horseshoe drag). It is therefore important to investigate the behaviour of the torque as a function of $b/h$.

For isothermal models, it has already been noted that there is a strong dependence on $b/h$, due to the $C^4$ scaling of the vortensity-related horseshoe drag \citep{drag}. Smaller values of $b/h$ give stronger corotation torques, but since $\xs$ is finite in the limit $b \rightarrow 0$ \citep{horse} the horseshoe drag is well-defined. The first term of the entropy-related horseshoe drag (see equation \ref{eqHSadi}) is proportional to $C^2h/b$, and therefore formally diverges for $b \rightarrow 0$. In our simple model, this stems from the stagnation point being located where the planet potential diverges. In practice, there will always be a finite distance from the stagnation point to the planet, and furthermore the planet potential should not diverge. For appropriate values of $d$ and therefore $\Phi_\mathrm{turn}$ equation \ref{eqHSadi} should still hold in the limit $b \rightarrow 0$. This limit is impossible to reach through numerical simulations, of course, since one always must resolve the gravitational potential in hydrodynamic simulations. 

We have varied $b/h$ between $0.2$ and $0.6$, and the results are displayed in Fig. \ref{figsoft}. Black lines indicate isothermal simulations for $\alpha=0$, which show a less negative torque for smaller softening parameters. Therefore, the horseshoe drag, being positive and stronger for smaller values of $b/h$, is more than able to compensate for the Lindblad torque, which is more negative for smaller $b/h$. This effect, due to $C$ being larger for smaller values of $b/h$, is nothing compared to the softening dependence of the entropy torque (see the grey curves in Fig. \ref{figsoft}). The total torque varies by an order of magnitude in this range of $b/h$, and this is including the more negative Lindblad torques for smaller $b/h$. Clearly, $b/h$ is a crucial parameter for the entropy-related torque.

The dotted lines in Fig. \ref{figsoft} have been obtained from equations similar to equation \ref{eqTtot}, for appropriate values of $C$ and $d$. We have found that a constant value for $d/b$ gives good results, i.e. the distance between the stagnation point  and the planet is a fixed number of softening lengths, irrespective of the value of $b/h$. The scaling of equation \ref{eqxs} breaks down, however, for $b/h<0.3$, and we have measured $C=1.26$ for $b/h=0.2$. This gives a good match to the simulation for the total torque. We note that in principle, one could come up with a more complicated functional form for $C(b/h)$, using the results from \cite{horse}. 

Although our numerical methods agree on the strong dependence of the torque on $b$, for small softening the differences become larger. This is illustrated in Fig. \ref{figsoftcomp}, where we compare the total torque for different softening parameters for RODEO and FARGO. For $b/h>0.4$ the agreement is very good, while for $b/h=0.2$ the difference is approximately $30\%$. Note that the source term in the vortensity equation becomes very strong at small softening, leading to a strong contribution from the entropy discontinuity at the outgoing separatrix. This is a very challenging situation for numerical methods, and it is not surprising that the differences between the methods are larger in this regime.

The solid curve in Fig. \ref{figsoftcomp} denotes equation \ref{eqTtotgen}, without the correction applied in Fig. \ref{figsoft} for the breakdown of the scaling of $C$ with $b/h$. Therefore, it predicts too large a torque at small softening (about $20\%$). Overall, however, the agreement is very good.
 
\begin{figure}
\centering
\resizebox{\hsize}{!}{\includegraphics[]{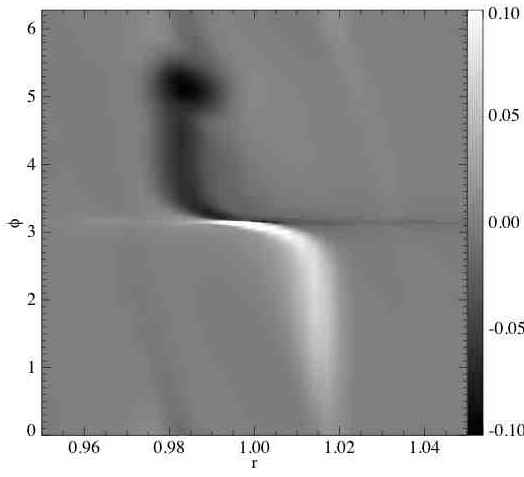}}
\caption{Perturbed vorticity $(\omega-\omega_0)/\omega_0$ due to the presence of a $q=1.26\cdot 10^{-5}$ planet n a disc with $h=0.05$, $\alpha=0$ and $\beta=1$ after $20$ orbits. The strong entropy gradient at the outgoing separatrix leads to the appearance of an anti-cyclonic vortex, visible as the circular shaped vorticity minimum at $(r,\varphi)=(0.985,5.2)$. Results were obtained with RODEO.}
\label{figvortex}
\end{figure}

\subsection{Vortex formation}
Vortices in protoplanetary discs can form as a result of the Rossby wave instability \citep{lovelace99}. This instability is usually discussed in the context of giant, gap opening planets \citep{li05,miguel07}, for which it was observed in most numerical codes for inviscid discs \citep{miguel06}. Although the low-mass planets discussed in this paper are not massive enough to significantly perturb the surface density,  the strong entropy gradients that exist at the outgoing separatrix can lead to vortex formation. This was observed in almost all simulations with a significant initial entropy gradient (see Fig. \ref{figvortex} for a typical example). \cite{baruteau08} also reported the appearance of a vortex in the same context. When the source term in the vorticity equation acts to decrease the vorticity, which is always true for one of the horseshoe legs, there is the possibility of forming an anti-cyclonic vortex. Since the formation of this vortex occurs after the turn, it does not affect the torque. This may not be the case when the vortex interacts with the planet when it reaches the opposite side; it may then affect the partial saturation of the corotation torque. This will be discussed in a forthcoming work. Note that a small kinematic viscosity is enough to kill the vortex before it reaches the opposite side of the planet. 

\begin{figure}
\centering
\resizebox{\hsize}{!}{\includegraphics[]{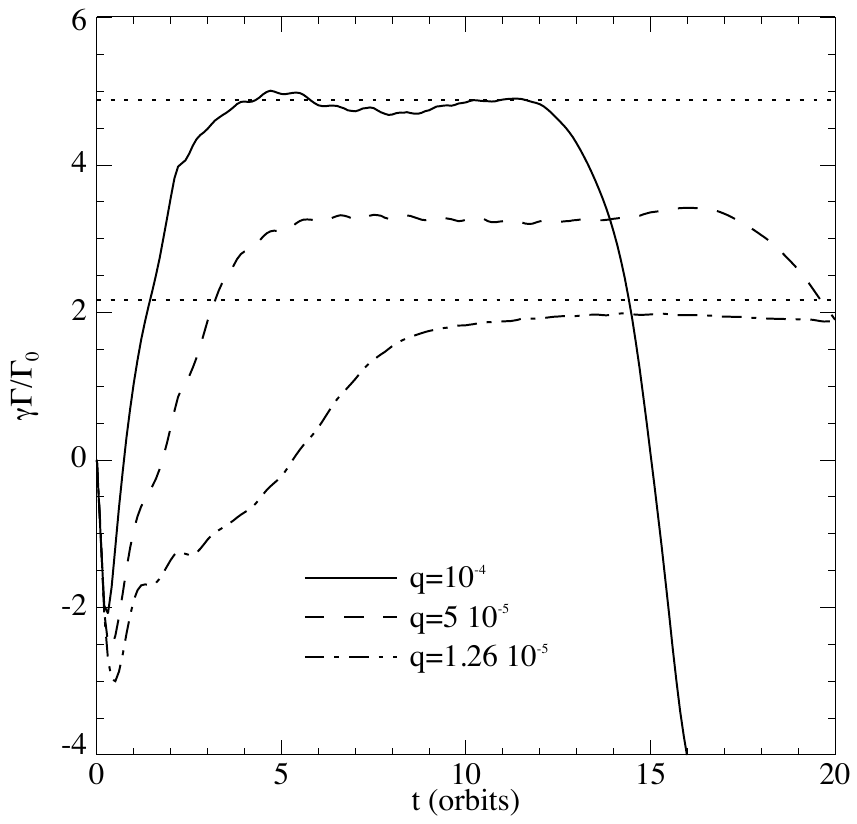}}
\caption{Total torque on planets of different mass (using $b/h=0.4$), embedded in an adiabatic $\gamma=5/3$ disc with $h=0.05$, $\alpha=0$ and $\beta=1$. The dotted lines denote equation \ref{eqTtot} for $C=1.3$ (top) and $C=1.1$ (bottom). Results were obtained with RODEO.}
\label{fighighmass}
\end{figure}

\subsection{Higher-mass planets}
\label{secHighmass}
Although the main focus of this paper in on low-mass planets, for which $\xs < h$, or, equivalently, $q < h^3$ \citep{horse}, there exist an interesting intermediate class of planets with masses of approximately $20$ - $50$ $\me$ that experience a boost in the corotation torque \citep{masset06}. It was shown in \cite{horse} that this is due to the fact that for $\xs > h$, Lindblad torques are less effective in affecting the shape of the horseshoe region. As a result, the width of the horseshoe region increases, and the stagnation points move closer to the planet. Both effects will enhance the horseshoe drag, in the isothermal case \citep{masset06}, but especially in the adiabatic case.

In Fig. \ref{fighighmass}, we show the total torque on planets of different mass. Note that since $\Gamma_0 \propto q^2$, curves for planets for which $\xs < h$ would fall on top of each other when the x-axis is rescaled in an appropriate way \citep{drag}. This is expected from linear theory, but this also holds when the non-linear horseshoe drag is incorporated, because $\xs \propto \sqrt{q}$ \citep{masset06,horse}. We found that for $q/h^3 > 0.2$, the boost as discussed in \cite{masset06} sets in. This can be seen in Fig. \ref{fighighmass} from the solid and dashed curves. For $q=10^{-4}=0.8h^3$ the maximum torque is reached. Beyond this mass, gap formation starts to play a role, lowering the mass and the opacity of the corotation region, which reduces the effect of the corotation torque. This was also observed in \cite{kley08}. 

The maximum torque in Fig. \ref{fighighmass}, achieved for $q=10^{-4}$, is in good agreement with equation \ref{eqTtot}, but with $C=1.3$. As mentioned in Sect. \ref{secTotaltorque}, since the stagnation point does not actually reach the location of the planet (see the bottom panels of Fig. \ref{figstream}), $C<1.3$, but this will be at least partly compensated by the decrease in $d$ compared to $d/b=\sqrt{13\gamma/4}$ as used in equation \ref{eqTtot}. The overall result is similar to taking $C=1.3$ (see Fig. \ref{fighighmass}), which is good enough for our purposes.  

We comment that a more general torque formula, valid for higher masses, would therefore have $C=C(b/h,q)$ and $d=d(q)$. A detailed analysis is beyond the scope of this paper, but we point out that the results of \cite{masset06} and \cite{horse} could be used to derive approximate functional forms for the mass dependence of $C$ and $d$.
  
\section{Discussion}
\label{secDisc}
In equation \ref{eqTtot}, we have presented a formula for the unsaturated torque on low-mass planets. This torque consists of the Lindblad torque, together with the barotropic and entropy-related horseshoe drag. The barotropic part of the horseshoe drag is due to material conserving its vortensity, and its expression is identical to the barotropic case. The entropy-related part of the horseshoe drag is exerted by density structures produced by material conserving its entropy, plus an additional component linked to the production of specific vorticity at the outgoing separatrices. \cite{masset09} recently studied the case with $\beta=0$ in detail, arguing that there is no contribution to the torque associated with a density response resulting from entropy advection. If we neglect this term in our simple model, and take $\bar v=3/2$, our expression for the density perturbation agrees with that in \cite{masset09}. We comment that the contribution of entropy advection to the total torque is small compared to that due to vortensity generation ($20-30 \%$, depending on the softening length), but keeping it gives better agreement with numerical simulations. 

Since we consider the unsaturated torque, equation \ref{eqTtot} should apply only in regions of the discs where thermal and viscous diffusion keep the corotation torque unsaturated. We will study saturation effects in a forthcoming work. Here, we just comment that it has been shown \citep{paardpap08,kley08} that when including viscosity as well as thermal diffusion (possibly through radiative effects) a sizable fraction of the unsaturated corotation torque can be sustained. 

We have worked in the 2D approximation throughout this paper. Although we have used a reasonable value for the gravitational softening parameter to mimic 3D averaging, a fully 3D model of the horseshoe region is required to capture possible effects due to vertical motions. The strong dependence of the torque on softening suggests that non-isothermal effects in 3D may be very strong, but it is important to keep in mind that the torque depends on the detailed flow structure around the planet, which may well be different in 3D.

We have neglected effects of any magnetic fields. It remains to be seen, for example, whether a fully turbulent disc \citep{nelson04} allows for horseshoe turns to occur. Self-gravity was also ignored. It was shown in \cite{pierens05} that self-gravity tends to make the wave torque slightly stronger due to a shift in the Lindblad resonances. This was confirmed numerically by \cite{baruteausg}, who also showed that the impact of self-gravity on the corotation torque is small. 

When calculating the torque on the planet, we have included all disc material. Tests have shown that it makes very little difference for these low-mass planets when a fraction of the Hill sphere is excluded. The situation is different for high-mass planets, for which a circumplanetary disc may appear. In those cases it is an issue which material should exert a torque on the planet \citep{crida09}. However, even the highest mass planets we consider do not show any evidence for a circumplanetary disc around the planet (see the bottom panels of Fig. \ref{figstream}), so this issue is of no concern here.

We have kept the planet on a fixed circular orbit. Therefore, we have neglected any distortion of the streamline topology due to the radial movement of the planet. This can have some impact on the corotation torque, especially for massive discs in which the planet migrates fast enough, under influence of the Type I torque discussed in this paper, so that $\dot \rp \sim \xs \op$. Then one may expect to see migration behaviour similar to Type III. This has not been considered so far. 

It is important to note that Type I migration will always be fast, unless the background disc is close to the zero-torque lines in Fig. \ref{figtorquetot}. The torque predicted by equation \ref{eqTtot} can indeed be much larger in magnitude than the linear, isothermal Type I torque (see equation \ref{eqTanaka3D}). Type I migration can be directed inward or outward, depending on the background entropy gradient. Outward migration is always limited, however, since inevitably the planet will enter a region of the disc where the opacity is low enough to make cooling efficient, pushing the planet back into the (locally) isothermal regime of inward migration. 

One scenario that permits slow migration only is the following. If the thermodynamic state of the inner disc is such that it permits outward Type I migration, there exists an equilibrium radius $r_\mathrm{e}$ where the torque is zero \citep{paard08}. A low-mass planet will then tend to migrate towards $r_\mathrm{e}$, either from the outer disc or from the inner disc. This radius $r_\mathrm{e}$ will move inward when the disc is losing mass, either by accretion onto the star or by evaporation, taking the planet along. This way, low-mass planets can migrate slowly (on a time scale comparable to the disc life time) towards the central star.   

\section{Conclusions}
\label{secCon}
We have presented a simple relation (equation \ref{eqTtot}) that governs the migration speed and direction for low-mass planets. Since we have considered unsaturated torques only, this law should apply in regions of the disc where thermal and viscous diffusion act to keep the corotation torque unsaturated. The total torque is found to strongly depend on the presence of a radial entropy gradient in the disc, with the possibility of outward migration in the case of outward decreasing entropy. 

\section*{Acknowledgements}
SJP acknowledges support from STFC in the form of a postdoctoral fellowship. This work was performed using the Darwin Supercomputer of the University of Cambridge High Performance Computing Service (http://www.hpc.cam.ac.uk), provided by Dell Inc. using Strategic Research Infrastructure Funding from the Higher Education Funding Council for England.

\bibliography{entropy.bib}

\begin{thebibliography}{}

\bibitem[\protect\citeauthoryear{{Alibert}, {Mordasini}, {Benz} \&
  {Winisdoerffer}}{{Alibert} et~al.}{2005}]{alibert05}
{Alibert} Y.,  {Mordasini} C.,  {Benz} W.,    {Winisdoerffer} C.,  2005, \aap,
  434, 343

\bibitem[\protect\citeauthoryear{{Artymowicz}}{{Artymowicz}}{1993}]{art93}
{Artymowicz} P.,  1993, \apj, 419, 155

\bibitem[\protect\citeauthoryear{{Baruteau} \& {Masset}}{{Baruteau} \&
  {Masset}}{2008a}]{baruteau08}
{Baruteau} C.,  {Masset} F.,  2008a, \apj, 672, 1054

\bibitem[\protect\citeauthoryear{{Baruteau} \& {Masset}}{{Baruteau} \&
  {Masset}}{2008b}]{baruteausg}
{Baruteau} C.,  {Masset} F.,  2008b, \apj, 678, 483

\bibitem[\protect\citeauthoryear{{Bate}, {Lubow}, {Ogilvie} \& {Miller}}{{Bate}
  et~al.}{2003}]{bate03}
{Bate} M.~R.,  {Lubow} S.~H.,  {Ogilvie} G.~I.,    {Miller} K.~A.,  2003,
  \mnras, 341, 213

\bibitem[\protect\citeauthoryear{{Boley}}{{Boley}}{2009}]{boley09}
{Boley} A.~C.,  2009, \apjl, 695, L53

\bibitem[\protect\citeauthoryear{{Boss}}{{Boss}}{1997}]{boss97}
{Boss} A.~P.,  1997, Science, 276, 1836

\bibitem[\protect\citeauthoryear{{Crida}, {Baruteau}, {Kley} \&
  {Masset}}{{Crida} et~al.}{2009}]{crida09}
{Crida} A.,  {Baruteau} C.,  {Kley} W.,    {Masset} F.,  2009, \aap, 502, 679

\bibitem[\protect\citeauthoryear{{Crida}, {Morbidelli} \& {Masset}}{{Crida}
  et~al.}{2006}]{crida06}
{Crida} A.,  {Morbidelli} A.,    {Masset} F.,  2006, Icarus, 181, 587

\bibitem[\protect\citeauthoryear{{D'Angelo}, {Henning} \& {Kley}}{{D'Angelo}
  et~al.}{2003}]{dangelo03}
{D'Angelo} G.,  {Henning} T.,    {Kley} W.,  2003, \apj, 599, 548

\bibitem[\protect\citeauthoryear{{D'Angelo}, {Kley} \& {Henning}}{{D'Angelo}
  et~al.}{2003}]{dangelo3D}
{D'Angelo} G.,  {Kley} W.,    {Henning} T.,  2003, \apj, 586, 540

\bibitem[\protect\citeauthoryear{{de Val-Borro}, {Artymowicz}, {D'Angelo} \&
  {Peplinski}}{{de Val-Borro} et~al.}{2007}]{miguel07}
{de Val-Borro} M.,  {Artymowicz} P.,  {D'Angelo} G.,    {Peplinski} A.,  2007,
  \aap, 471, 1043

\bibitem[\protect\citeauthoryear{{de Val-Borro}, {Edgar}, {Artymowicz},
  {Ciecielag}, {Cresswell} \& {D'Angelo}}{{de Val-Borro}
  et~al.}{2006}]{miguel06}
{de Val-Borro} M.,  {Edgar} R.~G.,  {Artymowicz} P.,  {Ciecielag} P.,
  {Cresswell} P.,    {D'Angelo} G.,  2006, \mnras, 370, 529

\bibitem[\protect\citeauthoryear{{Eulderink} \& {Mellema}}{{Eulderink} \&
  {Mellema}}{1995}]{eulderink95}
{Eulderink} F.,  {Mellema} G.,  1995, \aaps, 110, 587

\bibitem[\protect\citeauthoryear{{Goldreich} \& {Tremaine}}{{Goldreich} \&
  {Tremaine}}{1979}]{gt79}
{Goldreich} P.,  {Tremaine} S.,  1979, \apj, 233, 857

\bibitem[\protect\citeauthoryear{{Ida} \& {Lin}}{{Ida} \& {Lin}}{2008}]{ida08}
{Ida} S.,  {Lin} D.~N.~C.,  2008, \apj, 673, 487

\bibitem[\protect\citeauthoryear{{Jang-Condell} \& {Sasselov}}{{Jang-Condell}
  \& {Sasselov}}{2005}]{jang05}
{Jang-Condell} H.,  {Sasselov} D.~D.,  2005, \apj, 619, 1123

\bibitem[\protect\citeauthoryear{{Klahr} \& {Kley}}{{Klahr} \&
  {Kley}}{2006}]{klahr06}
{Klahr} H.,  {Kley} W.,  2006, \aap, 445, 747

\bibitem[\protect\citeauthoryear{{Kley}}{{Kley}}{1989}]{kley89}
{Kley} W.,  1989, \aap, 208, 98

\bibitem[\protect\citeauthoryear{{Kley}}{{Kley}}{1999}]{kley99}
{Kley} W.,  1999, \mnras, 303, 696

\bibitem[\protect\citeauthoryear{{Kley} \& {Crida}}{{Kley} \&
  {Crida}}{2008}]{kley08}
{Kley} W.,  {Crida} A.,  2008, \aap, 487, L9

\bibitem[\protect\citeauthoryear{{Korycansky} \& {Pollack}}{{Korycansky} \&
  {Pollack}}{1993}]{kory93}
{Korycansky} D.~G.,  {Pollack} J.~B.,  1993, Icarus, 102, 150

\bibitem[\protect\citeauthoryear{{Li}, {Li}, {Koller}, {Wendroff}, {Liska},
  {Orban}, {Liang} \& {Lin}}{{Li} et~al.}{2005}]{li05}
{Li} H.,  {Li} S.,  {Koller} J.,  {Wendroff} B.~B.,  {Liska} R.,  {Orban}
  C.~M.,  {Liang} E.~P.~T.,    {Lin} D.~N.~C.,  2005, \apj, 624, 1003

\bibitem[\protect\citeauthoryear{{Lin} \& {Papaloizou}}{{Lin} \&
  {Papaloizou}}{1986a}]{linpap86II}
{Lin} D.~N.~C.,  {Papaloizou} J.,  1986a, \apj, 307, 395

\bibitem[\protect\citeauthoryear{{Lin} \& {Papaloizou}}{{Lin} \&
  {Papaloizou}}{1986b}]{linpap86III}
{Lin} D.~N.~C.,  {Papaloizou} J.,  1986b, \apj, 309, 846

\bibitem[\protect\citeauthoryear{{Lovelace}, {Li}, {Colgate} \&
  {Nelson}}{{Lovelace} et~al.}{1999}]{lovelace99}
{Lovelace} R.~V.~E.,  {Li} H.,  {Colgate} S.~A.,    {Nelson} A.~F.,  1999,
  \apj, 513, 805

\bibitem[\protect\citeauthoryear{{Masset}}{{Masset}}{2000a}]{fargo}
{Masset} F.~S.,  2000a, \aaps, 141, 165

\bibitem[\protect\citeauthoryear{{Masset}}{{Masset}}{2000b}]{fargo2}
{Masset} F.~S.,  2000b, in {Garz{\'o}n} G.,  {Eiroa} C.,  {de Winter} D.,
  {Mahoney} T.~J.,  eds, Disks, Planetesimals, and Planets Vol.~219 of
  Astronomical Society of the Pacific Conference Series, {FARGO: A Fast
  Eulerian Transport Algorithm for Differentially Rotating Disks}.
p.~75

\bibitem[\protect\citeauthoryear{{Masset} \& {Casoli}}{{Masset} \&
  {Casoli}}{2009}]{masset09}
{Masset} F.~S.,  {Casoli} J.,  2009, \apj, 703, 857

\bibitem[\protect\citeauthoryear{{Masset}, {D'Angelo} \& {Kley}}{{Masset}
  et~al.}{2006}]{masset06}
{Masset} F.~S.,  {D'Angelo} G.,    {Kley} W.,  2006, \apj, 652, 730

\bibitem[\protect\citeauthoryear{{Masset} \& {Papaloizou}}{{Masset} \&
  {Papaloizou}}{2003}]{maspap03}
{Masset} F.~S.,  {Papaloizou} J.~C.~B.,  2003, \apj, 588, 494

\bibitem[\protect\citeauthoryear{{Menou} \& {Goodman}}{{Menou} \&
  {Goodman}}{2004}]{menou04}
{Menou} K.,  {Goodman} J.,  2004, \apj, 606, 520

\bibitem[\protect\citeauthoryear{{Mordasini}, {Alibert} \& {Benz}}{{Mordasini}
  et~al.}{2009}]{mordasini09}
{Mordasini} C.,  {Alibert} Y.,    {Benz} W.,  2009, ArXiv e-prints

\bibitem[\protect\citeauthoryear{{Nelson} \& {Papaloizou}}{{Nelson} \&
  {Papaloizou}}{2004}]{nelson04}
{Nelson} R.~P.,  {Papaloizou} J.~C.~B.,  2004, \mnras, 350, 849

\bibitem[\protect\citeauthoryear{{Nelson}, {Papaloizou}, {Masset} \&
  {Kley}}{{Nelson} et~al.}{2000}]{nelson00}
{Nelson} R.~P.,  {Papaloizou} J.~C.~B.,  {Masset} F.,    {Kley} W.,  2000,
  \mnras, 318, 18

\bibitem[\protect\citeauthoryear{{Paardekooper} \& {Mellema}}{{Paardekooper} \&
  {Mellema}}{2006a}]{paard06}
{Paardekooper} S.-J.,  {Mellema} G.,  2006a, \aap, 459, L17

\bibitem[\protect\citeauthoryear{{Paardekooper} \& {Mellema}}{{Paardekooper} \&
  {Mellema}}{2006b}]{rodeo}
{Paardekooper} S.-J.,  {Mellema} G.,  2006b, \aap, 450, 1203

\bibitem[\protect\citeauthoryear{{Paardekooper} \& {Mellema}}{{Paardekooper} \&
  {Mellema}}{2008}]{paard08}
{Paardekooper} S.-J.,  {Mellema} G.,  2008, \aap, 478, 245

\bibitem[\protect\citeauthoryear{{Paardekooper} \& {Papaloizou}}{{Paardekooper}
  \& {Papaloizou}}{2008}]{paardpap08}
{Paardekooper} S.-J.,  {Papaloizou} J.~C.~B.,  2008, \aap, 485, 877

\bibitem[\protect\citeauthoryear{{Paardekooper} \& {Papaloizou}}{{Paardekooper}
  \& {Papaloizou}}{2009a}]{drag}
{Paardekooper} S.-J.,  {Papaloizou} J.~C.~B.,  2009a, \mnras, 394, 2283

\bibitem[\protect\citeauthoryear{{Paardekooper} \& {Papaloizou}}{{Paardekooper}
  \& {Papaloizou}}{2009b}]{horse}
{Paardekooper} S.-J.,  {Papaloizou} J.~C.~B.,  2009b, \mnras, 394, 2297

\bibitem[\protect\citeauthoryear{{Pepli{\'n}ski}}{{Pepli{\'n}ski}}{2008}]{adam%
thesis}
{Pepli{\'n}ski} A.,  2008, PhD thesis, Department of Astronomy, Stockholm
  University, Stockholm, Sweden

\bibitem[\protect\citeauthoryear{{Pepli{\'n}ski}, {Artymowicz} \&
  {Mellema}}{{Pepli{\'n}ski} et~al.}{2008a}]{adamin}
{Pepli{\'n}ski} A.,  {Artymowicz} P.,    {Mellema} G.,  2008a, \mnras, 386, 179

\bibitem[\protect\citeauthoryear{{Pepli{\'n}ski}, {Artymowicz} \&
  {Mellema}}{{Pepli{\'n}ski} et~al.}{2008b}]{adamout}
{Pepli{\'n}ski} A.,  {Artymowicz} P.,    {Mellema} G.,  2008b, \mnras, 387,
  1063

\bibitem[\protect\citeauthoryear{{Pierens} \& {Hur{\'e}}}{{Pierens} \&
  {Hur{\'e}}}{2005}]{pierens05}
{Pierens} A.,  {Hur{\'e}} J.-M.,  2005, \aap, 433, L37

\bibitem[\protect\citeauthoryear{{Pollack}, {Hubickyj}, {Bodenheimer},
  {Lissauer}, {Podolak} \& {Greenzweig}}{{Pollack} et~al.}{1996}]{pollack96}
{Pollack} J.~B.,  {Hubickyj} O.,  {Bodenheimer} P.,  {Lissauer} J.~J.,
  {Podolak} M.,    {Greenzweig} Y.,  1996, Icarus, 124, 62

\bibitem[\protect\citeauthoryear{{Tanaka}, {Takeuchi} \& {Ward}}{{Tanaka}
  et~al.}{2002}]{tanaka02}
{Tanaka} H.,  {Takeuchi} T.,    {Ward} W.~R.,  2002, \apj, 565, 1257

\bibitem[\protect\citeauthoryear{{Terquem}}{{Terquem}}{2003}]{terquem03}
{Terquem} C.~E.~J.~M.~L.~J.,  2003, \mnras, 341, 1157

\bibitem[\protect\citeauthoryear{{van Leer}}{{van Leer}}{1977}]{vanLeer77}
{van Leer} B.,  1977, Journal of Computational Physics, 23, 276

\bibitem[\protect\citeauthoryear{{Ward}}{{Ward}}{1991}]{ward91}
{Ward} W.~R.,  1991, in Lunar and Planetary Institute Conference Abstracts
  {Horsehoe Orbit Drag}.
p.~1463

\bibitem[\protect\citeauthoryear{{Ward}}{{Ward}}{1997}]{ward97}
{Ward} W.~R.,  1997, Icarus, 126, 261

\bibitem[\protect\citeauthoryear{{Weidenschilling}}{{Weidenschilling}}{1977}]{%
weiden77}
{Weidenschilling} S.~J.,  1977, \mnras, 180, 57

\end{thebibliography}

\label{lastpage}
\end{document}